\documentclass[final,12pt,times,authoryear]{elsarticle}
\usepackage{multicol}
\usepackage{bm}
\usepackage{lipsum}
\usepackage{mathtools}
\usepackage{amssymb}
\usepackage{amsmath}
\usepackage{eqnarray}
\usepackage{graphicx}
\usepackage{natbib}
\usepackage{multirow} 
\usepackage{threeparttable}
\usepackage{tabularx} 
\usepackage{dblfloatfix}
\usepackage{blindtext}
\usepackage{color}
\usepackage{subfig}
\usepackage{setspace}
\usepackage{setspace}
\begin{document}
\begin{frontmatter}
\title{Bayesian Inference for Johnson's SB and Weibull distributions}
\author{Mahdi Teimouri}
\address{Department of Mathematics and Statistics, Faculty of Science and Engineering, Gonbad Kavous University, Gonbad Kavous, Iran, Email: teimouri@aut.ac.ir}
\begin{abstract}
The four-parameter Johnson's SB (JSB) and three-parameter Weibull distributions have received much attention in the field of forestry for characterizing diameters at breast height (DBH). In this work, we suggest the Bayesian method for estimating parameters of the JBS distribution. The maximum likelihood approach  uses iterative methods such as Newton-Raphson (NR) algorithm for maximizing the logarithm of the likelihood function. But there is no guarantee that the NR method converges. This fact that the NR method for estimating the parameters of the JSB distribution sometimes fails to converge was verified through simulation in this study. Further, it was shown that the Bayesian estimators presented in this work were robust with respect to the initial values and estimate the parameters of the JSB distribution efficiently. The performance of the JSB and three-parameter Weibull distributions was compared in a Bayesian paradigm when these models were fitted to DBH data of three plots that randomly selected from a study established in 107 plots of mixed-age ponderosa pine ({\it{Pinus ponderosa}} Dougl. ex Laws.) with scattered western junipers at the Malheur National Forest in south end of the Blue Mountains near Burns, Oregon, USA. Bayesian paradigm demonstrated that JBS was superior model than the three-parameter Weibull for characterizing the DBH distribution when these models were fitted to the DBH data of the three plots.
\end{abstract}
\end{frontmatter}
\section{Introduction}
Statistical modelling for the distribution of the diameter at breast height (DBH) is becoming increasingly popular in order to characterizing the forest height structure, forest dynamics, and comparing the forest stands \citep{gorgoso2007modelling, mateus2011modelling, ozccelik2016modeling}. The statistical characterization or modelling of the DBH distribution has a long history in both managed and natural forest stands. Among all statistical models, the desired is that shows more flexibility, i.e., capturing well the DBH distribution. This is because different types of the forest stands show different shape for DBH distribution. For example, two main types of the forest stands include the even-aged that usually are unimodal (one peak) and roughly symmetric and uneven-aged that whose DBH distributions often have a reverse-$J$ shape. Among all statistical distributions, the Johnson's SB (JSB) and Weibull have received much attention in the context forest management. Numerous efforts have been made in the literature for modelling the tree's DBH among them are \citep{Bailey1973quantifying, maltamo1995comparison, maltamo2000comparison, pretzsch2009forest, zhang2010compatibility} for two- or three-parameter Weibull distribution, \citep{fonseca2009describing,hafley1985bivariate,kiviste2003diameter,kudus1999nonlinear,marto2009computer,
mateus2011modelling,ozccelik2016modeling,parresol2003recovering,rennolls2005new,zhou1996comparison} for JSB distribution, and  \citep{gorgoso2012comparison,hafley1977statistical,palahi2007comparison,zhang2003comparison} for both of them. In statistical modelling of DBH, precision of the parameter estimator plays a crucial role in the forest planning and management \citep{gorgoso2007modelling, mateus2011modelling, ozccelik2016modeling}. 
\par The maximum likelihood (ML) approach, as the most popular estimation method, is obtained with the aid of mathematical optimization tools. These tools maximize the logarithm of the likelihood (log-likelihood) function using iterative algorithm such as Newton-Raphson (NR) and so need the initial values. If the initial values are far away from the true parameter (which is where the log-likelihood function reaches its global maximum), or when the log-likelihood function at the initial values becomes large, then there is no guarantee that the NR method will converge. This means that the ML approach is sensitive to the initial values and really can be considered as a weakness for the ML approach. Also, the ML approach may break down when the regularity conditions fail to exist. The above criticisms may happen when one is interested in estimating the parameters of JSB and three-parameter Weibull distributions. Other methods such as moment-based estimators are not as efficient as the ML approach. For example, moment-based estimators of the three-parameter Weibull distribution overcome the weaknesses in the ML approach, but their existence, uniqueness, and consistency are still open questions \citep{nagatsuka2013consistent} or in the case of JBS distribution, the moment-based estimator are not as efficient as regression-type estimators \citep{scolforo2003sb}. 
The aim of this study is to derive the Bayesian estimators for the parameters of the JSB and three-parameter Weibull distributions, that to the best of our knowledge, the Bayesian estimators of the parameters of JSB distribution have never been used in the forestry literature for modeling DBH distributions. This paper is organized as follows. In what follows we give some preliminaries. The Bayesian paradigm for the JSB and Weibull distributions are given in Section \ref{sec2}. Section \ref{sec3} is devoted to the materials and methods. We give the results and discussion in Sections \ref{sec4} and \ref{sec5}, respectively. We conclude the paper in Section \ref{sec6}.
\par
In the following, we give some preliminaries. 
\subsection{The family of JSB and three-parameter Weibull distributions}
The probability density function (pdf) and cumulative distribution function (cdf) of the Johnson SB (JSB) are given, respectively, by  \citep{johnson1949systems,norman1994continuous}:
\begin{equation}\label{BSpdf}
g_J\bigl(x\big|\Theta\bigr) = \frac {\delta \lambda}{\sqrt{2\pi}(x-\xi)(\lambda+\xi-x)}\exp\Biggl\{ -\frac{1}{2}\Bigg[\gamma+\delta\log \biggl(\frac{x-\xi}{\lambda+\xi-x}\biggr) \Bigg]^2\Biggr\},
\end{equation}
and
\begin{equation}\label{BScdf}
G_J\bigl(x\big|\Theta\bigr) =\int_{\xi}^{x}g_J\bigl(y\big|\Theta\bigr) dy,
\end{equation}
where $\Theta=(\delta,\gamma,\lambda,\xi)^T$, $\xi<x<\lambda+\xi$, $\delta>0$, $\lambda> 0$, $-\infty<\gamma<\infty$, and $-\infty<\xi<\infty$. As it is seen, the cdf of the JBS distribution has no closed-form expression. The pdf and cdf of three-parameter Weibull distribution are given, respectively, by
\begin{equation}\label{Wpdf}
g_W\bigl(x\big|\Theta\bigr) =\frac{\alpha}{\beta} \biggl(\frac {x-\mu}{\beta}\biggr)^{\alpha-1}\exp\Biggl\{ -\biggl( \frac {x-\mu}{\beta}\biggr)^\alpha \Biggr\},
\end{equation}
and
\begin{equation}\label{Wcdf}
G_W\bigl(x\big|\Theta\bigr) = 1-\exp\Biggl\{ -\biggl( \frac {x-\mu}{\beta}\biggr)^\alpha \Biggr\},
\end{equation} 
where $\Theta=(\alpha,\beta,\mu)^T$, $\mu<x$, $\alpha>0$, and $\beta>0$. Now, $\alpha$, $\beta$, and $\mu$ are the shape, scale, and location parameters, respectively.
\subsection{Bayes theorem}
In the Bayesian framework, we assume that the unknown parameter vector $\Theta$ follows a distribution with pdf $\pi(\Theta)$. Using information available in random observations $\boldsymbol{x}=(x_1,\dots,x_n)^T$, a revision will be made on knowledge about $p(\Theta)$ using the well-known Bayes' theorem as $\pi(\Theta|\boldsymbol{x})$. We have:
\begin{equation*}
\pi(\Theta|\boldsymbol{x})=\frac{g(\boldsymbol{x}| \Theta)\pi(\Theta)}{g(\boldsymbol{x})}
\end{equation*}
The expression $\pi(\Theta)$ and $\pi(\Theta|\boldsymbol{x})$ are known in the literature as prior pdf and posterior pdf of $\Theta$, respectively. Here, $g(\boldsymbol{x})$ is normalizing constant and so the Bayes' theorem can be written as
 \begin{equation}\label{post}
\pi(\Theta|\boldsymbol{x})\propto g(\boldsymbol{x}| \Theta)\pi(\Theta)
\end{equation}
\subsection{The NR algorithm for JSB distribution}
As previously mentioned, the NR algorithm may fail to converge. Unfortunately, this happens when finding the ML estimators of the JSB distribution is desired. We performed a simulation study to prove our claim. So, a number of 10,000 samples with different sizes including 20, 50, 100, 250, 500, 1000, and 5,000 were simulated from the JSB distribution with pdf given in (\ref{BSpdf}) and then we obtained the ML estimators of the parameters using the command \texttt{optim(.)} in R \citep{team2018r} environment. In each of 10,000 runs, the parameters $\delta$, $\gamma$, $\lambda$, and $\xi$ were generated from uniform distribution (0.05,10), (-20,20), (1,100), and (-50,50), respectively. While $x^{(i)}_{(1)}$ and $x^{(i)}_{(1)}$ denote, respectively, the smallest and largest values of the $i$-th generated sample, for $i=1,\dots,10000$, the initial values of $\delta$, $\gamma$, $\lambda$, and $\xi$ were generated from uniform distribution (0.05,10), (-20,20), $\Bigl(x^{(i)}_{(n)}-x^{(i)}_{(1)},100\Bigr)$, and $\Bigl(-50,x^{(i)}_{(1)}\Bigr)$, respectively. The results of simulation are given in Table \ref{tab1}. As it is seen, percentage of failed attempts to reach convergence through the NR algorithm is considerable (say, on the average 32\%).
\subsection{Gibbs sampler}
Due to the complicate nature of the posterior pdf $\pi(\Theta|\boldsymbol{x})$, we have to sample from the posterior pdf and then the Bayesian estimators are the ergodic average of the generated sample. In practice, exploitation of $\pi(\Theta|\boldsymbol{x})$ needs the use of Markov chain Monte Carlo (MCMC) techniques. The Gibbs sampler is one of the Markov chain Monte Carlo (MCMC) techniques that enables us to sample from full conditional pdf, i.e., the pdf of each element of the parameter vector given the other elements and observed data $\boldsymbol{x}=(x_1,\dots,x_n)^T$. Suppose that we have a statistical distribution with unknown parameter vector $\Theta=(\theta_1,\theta_2\dots,\theta_k)^T$ and observed $\boldsymbol{x}$. Consider $\Theta^{(0)}=\bigl(\theta^{(0)}_1,\theta^{(0)}_2\dots,\theta^{(0)}_k\bigr)^T$ as the initial values. In order to implement the Gibbs sampler technique, we generate $\theta^{(1)}_1$ from the full conditional pdf $\pi\bigl(\theta_{1}\big|\theta^{(0)}_2,\theta^{(0)}_3,\dots,\theta^{(0)}_k,\boldsymbol{x}\bigr)$, $\theta^{(1)}_2$ from the full conditional pdf $\pi\bigl(\theta_{2}\big|\theta^{(1)}_1,\theta^{(0)}_3,\dots,\theta^{(0)}_k,\boldsymbol{x}\bigr)$, and so on up to $\theta^{(1)}_k$ from the full conditional pdf $\pi\bigl(\theta_{k}\big|\theta^{(1)}_1,\theta^{(1)}_2,\dots,\theta^{(1)}_{k-1},\boldsymbol{x}\bigr)$. In the first iteration we obtain the sample $\Theta^{(1)}=\bigl(\theta^{(1)}_1,\theta^{(1)}_2\dots,\theta^{(1)}_k\bigr)^T$. Under mild regularity conditions \citep{roberts1994simple}, after a sufficiently large number of iterations, say $N$, the ergodic average of the Markov chain yields a consistent estimator of $\Theta$. 
\subsection{Metropolis-Hastings algorithm}
The Metropolis-Hastings (MH) algorithm, is an MCMC technique and efficient method for drawing samples from a given posterior distribution. Assume that we want to simulate sample from pdf given by (\ref{post}). Firstly, we need to choose a proposal distribution $q(.|.)$, that changes the location of the chain at each iteration of the algorithm. The proposal distribution is arbitrary and can be chosen so that is easy to simulate from. Secondly, we follow the steps given by the following.
\begin{itemize}
\item Choose initial value $\theta^{(0)}$ and  set i=1;
\begin{enumerate}
\item Sample $\theta^{*}$ from $q\bigl(.\big|\theta^{(i-1)}\bigr)$;
\item Set $\theta^{(i)}=\theta^{*}$ with probability
\begin{equation*}
\eta=\min \left\{1,\frac{\pi\bigl(\theta^{*}\big|\boldsymbol{x}\bigr)q\bigl(\theta^{(i-1)}\big|\theta^{*}\bigr)}{\pi\bigl(\theta^{(i-1)}\big|\boldsymbol{x}\bigr)q\bigl(\theta_{*}\big|\theta^{(i-1)}\bigr)}\right\},
\end{equation*}
otherwise $\theta^{(i)}=\theta^{(i-1)}$;
\item Set i=i+1 and go to step 1;
\end{enumerate}
\item Stop the algorithm if $i=N$ ($N$ is a sufficiently large integer value). 
\end{itemize}
Since realization generated at each iteration is used to generate sample at next step, the chain constitutes a correlated stochastic process, but after $N$ numbers of generations, we hope that the chain produces uncorrelated samples and converges to the target distribution $\pi(\theta|\boldsymbol{x})$ as desired.   
\subsection{Adaptive rejection sampling}
For a given sample $\boldsymbol{x}=(x_1,\dots,x_n)^T$, suppose we are interested in sampling from posterior pdf $\pi(\theta|\boldsymbol{x})$. The well-known accept-reject sampling needs a suitable upper bound, called $M$ that satisfies in the following inequality.
\begin{equation*}
\underset{\theta}{\operatorname{sup}}\frac{\pi(\theta|\boldsymbol{x})}{g(\theta)} \leq M,
\end{equation*}
where $g(\theta)$ is an arbitrary pdf that is easy to sample from and its support includes the support of $\pi(\theta|\boldsymbol{x})$. If there exists a suitable choice for $M$, then the accept-reject sampling is efficient otherwise we refer to another Monte Carlo simulation technique, called adaptive rejection sampling (ARS) algorithm. The ARS algorithm is used to simulate realization when posterior pdf is log-concave (i. e., the second derivative of $\pi(\theta|\boldsymbol{x})$ with respect to $\theta$ is negative). In such a case, the ARS algorithm developed by \cite{gilks1992adaptive} is highly efficient.
\section{Bayesian analysis}\label{sec2}
Here, we give the Bayesian paradigm for the JSB and Wribull distributions, respectively.
\subsection{Gibbs sampler for JSB distribution}
For Bayesian inference of the JSB distribution parameters, we assume that all four priors are statistically independent and so the full Bayesian model (joint posterior pdf) up to proportionality becomes
\begin{align}
\pi(\Theta|\boldsymbol{x}) &\propto g_{J}(\boldsymbol{x}|\Theta)\pi(\delta, \gamma, \lambda, \xi)\nonumber\\
&=\Pi_{i=1}^{n} g_{J}(x_i|\Theta)\pi(\delta)\pi(\gamma)\pi(\lambda)\pi(\xi)\nonumber\\
&= \frac{\delta^n\lambda^n}{(2\pi)^{\frac{n}{2}}\Pi_{i=1}^{n}(x_i-\xi)(\lambda+\xi-x_i)}
\exp\left\{-\frac{1}{2}\sum_{i=1}^{n}\Bigg[\gamma+\delta \log\biggl(\frac{x_i-\xi}{\lambda+\xi-x_i}\biggr)\Bigg]^2\right\}
\pi(\delta)\pi(\gamma)\pi(\lambda)\pi(\xi).
\end{align}
We note that we produce the Gibbs sampler directly from the full conditionals by choosing improper priors, i.e., bypassing the propriety of the posterior. This is due to the fact that the full conditionals are well-defined and the Bayesian model under study is enough complex \citep{robert2010introducing}. Details for generating from full conditionals of $\delta$, $\gamma$, $\lambda$, and $\xi$ are given \ref{apa}, \ref{apb}, \ref{apc}, and \ref{apd}, respectively.
\subsection{Gibbs sampler for three-parameter Weibull distribution}
The Bayesian paradigm for the three-parameter Weibull distribution originally was developed by \cite{smith1987comparison} and \cite{green1994bayesian}. Here, we give a slightly different version of the Bayesian paradigm developed by \cite{green1994bayesian}. We mention that the only difference occurs in updating the location parameter at each iteration if the chain. Assume that $\boldsymbol{x}=(x_1,\dots,x_n)^T$ denotes the vector of $n$ independent observations each follows distribution with pdf given in (\ref{Wpdf}). We consider the Jeffreys' prior \citep{Jeffreys} for $\alpha$ and $\beta$, i.e., $\pi(\alpha)\propto1/\alpha$ and $\pi(\beta)\propto1/\beta$. Also, we allow the prior for $\mu$ to be uniformly over ${{\mathbb{R}}}$. So, the full Bayesian model is given by 
\begin{align}
\pi(\Theta|\boldsymbol{x}) &\propto g_{W}(\boldsymbol{x}|\Theta)\pi(\alpha, \beta, \mu)\nonumber\\
&=\Pi_{i=1}^{n} g_{W}(x_i|\Theta)\pi(\alpha)\pi(\beta)\pi(\mu)\nonumber\\
&= \frac{\alpha^{n-1}}{\beta^{n+1}} \Pi_{i=1}^{n}\biggl(\frac{x_i-\mu}{\beta}\biggr)^{\alpha-1}
\exp\left\{-\sum_{i=1}^{n}\biggl(\frac{x_i-\mu}{\beta}\biggr)^{\alpha}\right\}.
\end{align}
The full conditionals of $\alpha$, $\beta$, and $\mu$ are (up to proportionality) given by
\begin{align}
\pi(\alpha|\beta,\mu,\boldsymbol{x}) &\propto \alpha^{n-1} \Pi_{i=1}^{n}\biggl(\frac{x_i-\mu}{\beta}\biggr)^{\alpha-1} \exp\left\{-\sum_{i=1}^{n}\biggl(\frac{x_i-\mu}{\beta}\biggr)^{\alpha}\right\}\label{alphapost},\\
\pi(\beta|\alpha,\mu,\boldsymbol{x}) &\propto \beta^{-n\alpha-1} \exp\left\{-\sum_{i=1}^{n}\biggl(\frac{x_i-\mu}{\beta}\biggr)^{\alpha}\right\}\label{betapost},\\
\pi(\mu|\alpha,\beta,\boldsymbol{x}) &\propto \Pi_{i=1}^{n}\bigl(x_i-\mu\bigr)^{\alpha-1} \exp\left\{-\sum_{i=1}^{n}\biggl(\frac{x_i-\mu}{\beta}\biggr)^{\alpha}\right\}\label{mupost}.
\end{align}
As pointed out by \cite{green1994bayesian}, since $\pi(\alpha|\beta,\mu,\boldsymbol{x})$ given in (\ref{alphapost}) is log-concave, the ARS algorithm is highly efficient technique for simulating from this full conditional.
Assuming that we are currently at the $t$-th iteration of the chain and we have just obtained $\alpha^{(t+1)}$ by simulating from full conditional given in (\ref{alphapost}). In order to generate from the full conditional $\pi\bigl(\beta\big|\alpha^{(t+1)},\mu^{(t)},\boldsymbol{x}\bigr)$ given in (\ref{betapost}), it suffices to simulate a realization from gamma distribution with shape parameter $n$, say $z$, and then update $\beta^{(t)}$ as $\beta^{(t+1)}$ using the relation 
\begin{align*}
\beta^{(t+1)}=\Biggl(\frac{\sum_{i=1}^{n}\bigl(x_i-\mu^{(t)}\bigr)^{\alpha^{(t+1)}}}{z}\Biggr)^{\frac{1}{\alpha^{(t+1)}}}.
\end{align*}
For the location parameter with full conditional given in (\ref{mupost}), we do not follow the accept-reject sampling method proposed by \cite{green1994bayesian}. Our study revealed that when $\alpha$ is small (say $\alpha<2$) the accept-reject sampling method does not work efficiently (the chin takes too much time for updating). Instead, we use the MH algorithm for generating from full conditional of $\mu$. For this purpose, we use the uniform distribution on $\bigl(x_{(1)}-\beta,x_{(1)}\bigr)$ as the proposal. The steps of the MH algorithm are given by the following.
\begin{enumerate}
\item Suppose we are currently at the $t$-th iteration of the chain. Set $i=1$ and choose the initial value as $\mu^{(0)}=x_{(1)}-1/n$;
\item Sample $\mu^{*}$ from proposal distribution with pdf $q\bigl(\mu)=1/\beta^{(t+1)}$, for $x_{(1)}-\beta^{(t+1)}<\mu<x_{(1)}$;
\item Compute $\eta$ as
\begin{equation*}
\eta=\min \left\{1,\frac{\pi\bigl(\mu^{*}\big|\alpha^{(t+1)}, \beta^{(t+1)}, \boldsymbol{x}\bigr)}{\pi\bigl(\mu_{(i-1)}\big|\alpha^{(t+1)}, \beta^{(t+1)}, \boldsymbol{x}\bigr)}\right\}.
\end{equation*}
\item Generate an uniform random variable on $(0,1)$, say $u$. If $u<\eta$, then $\xi^{(i)}=\xi^{*}$, otherwise $\xi^{(i)}=\xi^{(i-1)}$;
\item If $i=N$, then go to the next step. Otherwise set $i=i+1$ and go to step 2;
\item Accept $\mu_{(N)}$ as a generation form pdf $\pi\bigl(\mu\big|\alpha^{(t+1)}, \beta^{(t+1)}, \boldsymbol{x}\bigr)$, i.e., $\mu^{(t+1)}=\mu_{(N)}$ and stop the MH algorithm. 
\end{enumerate}
We note that the MH algorithm adopted for simulating from full conditional of $\mu$ is faster than the accept-reject sampling method proposed by \cite{green1994bayesian} when $\alpha<2$. Also, by choosing a uniform proposal, we allow the location parameter to vary over real line. This feature of our proposed prior for location parameter will appeal to a wide range of study fields in which modelling data through three-parameter Weibull distribution with negative location parameter occurs frequently.
\section{Materials and methods}\label{sec3}
\subsection{Materials}
Since our work motivated by the widespread use and application of the statistical distributions in the forest management, we conducted a study in the context of the forestry. A study was established in 107 plots of mixed-age ponderosa pine ({\it{Pinus ponderosa}} Dougl. ex Laws.) with scattered western junipers (\emph{Juniperus occidentalis} Hook.) that located in Malheur National Forest in south end of the Blue Mountains near Burns, Oregon, USA \cite{Kerns}. These data include tree height, diameter, and growth for a prescribed burning study with unburned controls. Of these variables, we only used the DBH (measured at a height of 1.3 m) of all live trees in three randomly selected plots (plots 9, 73, and 81) each of size 0.08 ha for statistical validation of the Bayesian approach. The plots summary statistics are given in Table \ref{tab2}. 
\subsubsection{Methods}
\begin{figure}
\includegraphics[width=85mm,height=85mm]{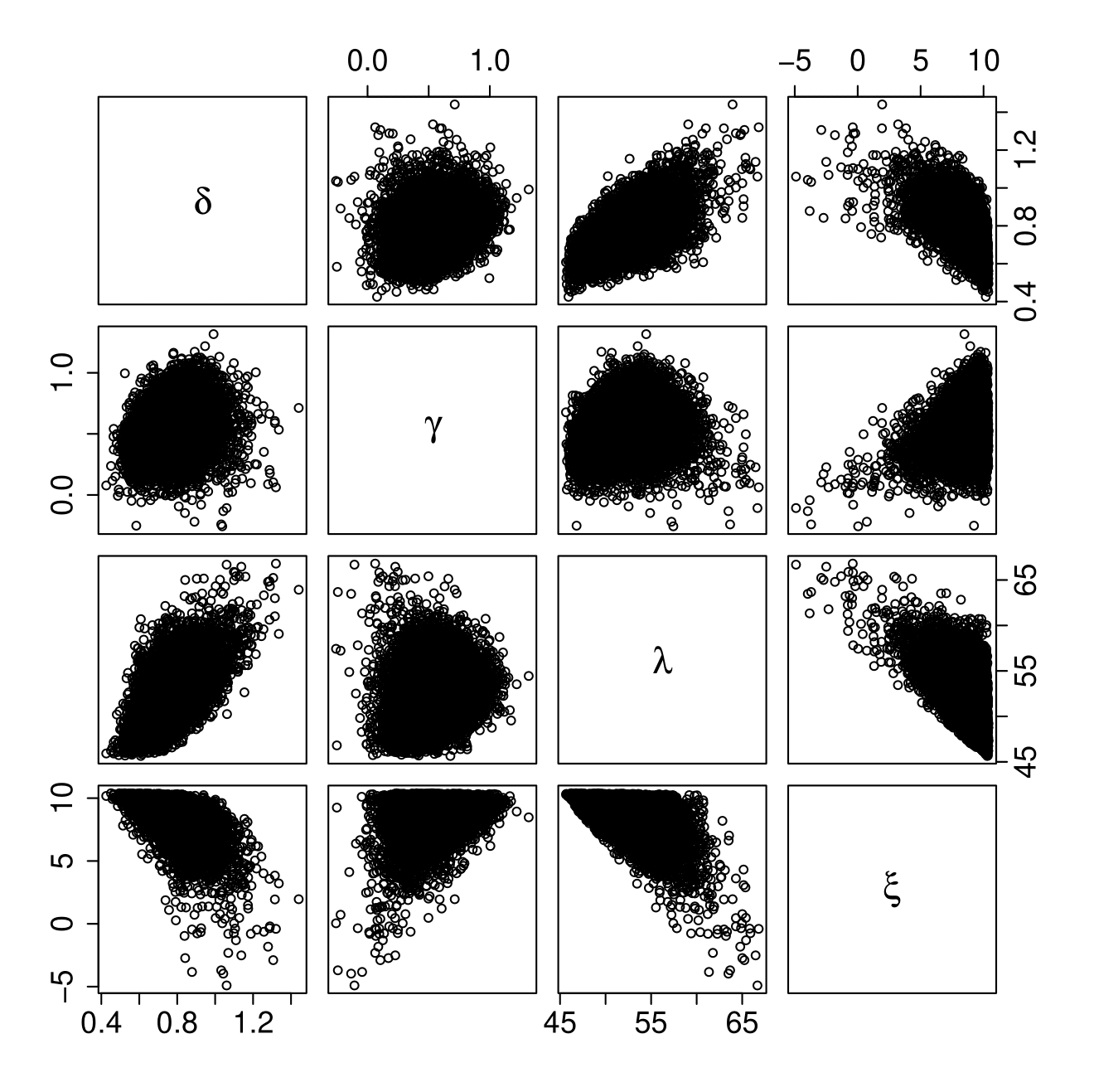}
\includegraphics[width=85mm,height=85mm]{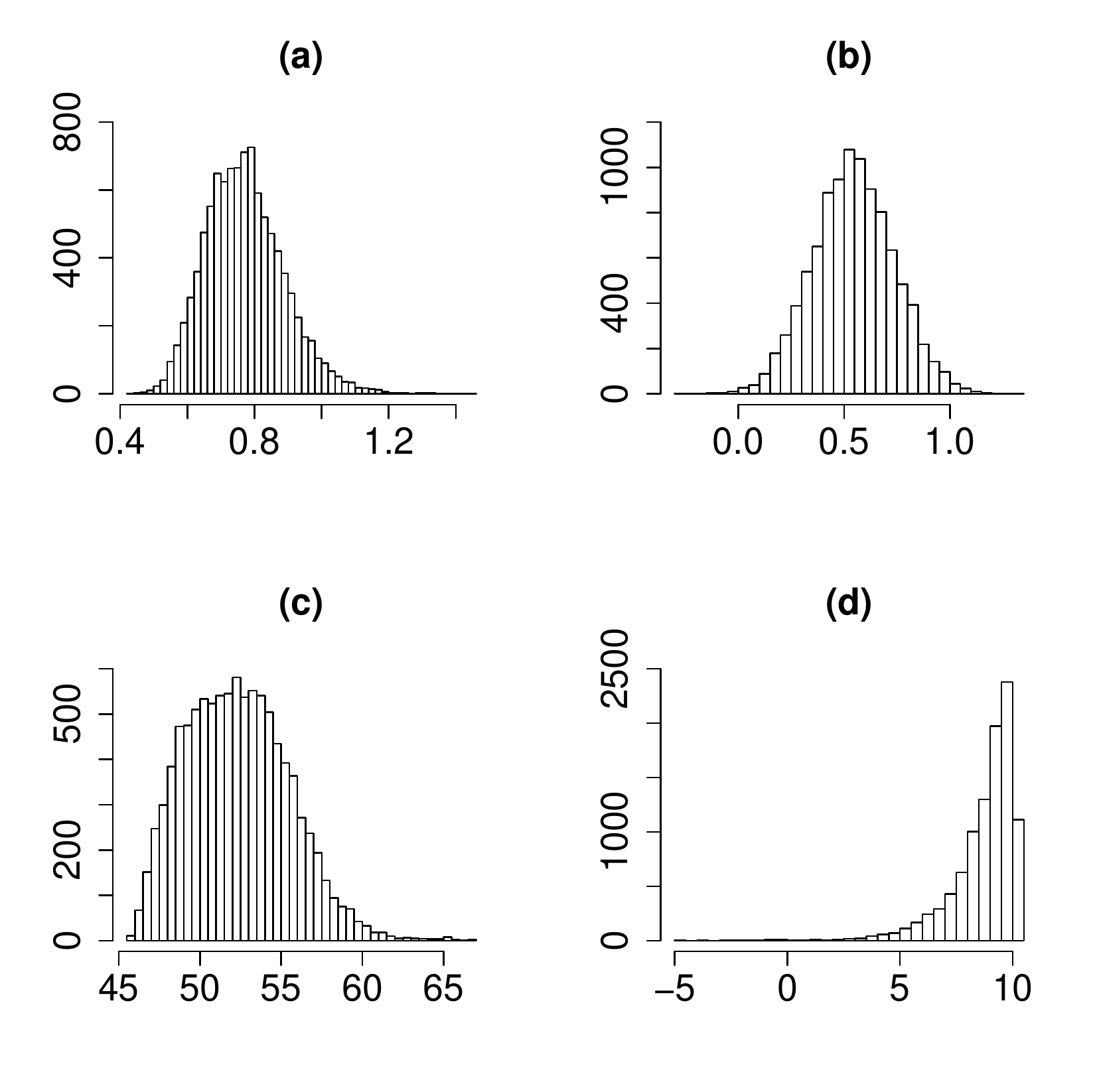}
\caption{Left-hand side: pairwise scatterplots of trimmed output of the Gibbs sampler for estimation parameters of the JSB distribution fitted to the DBH observations in plot 9. These outputs suggest that there is little dependence between $\delta$ and $\lambda$. Right-hand side: histograms of the full conditionals (a) $\pi(\delta|\gamma,\lambda,\xi,\boldsymbol{x})$, (b) $\pi(\gamma|\delta,\lambda,\xi,\boldsymbol{x})$, (c) $\pi(\lambda|\delta,\gamma,\xi,\boldsymbol{x})$, and (d) $\pi(\xi|\delta,\gamma,\lambda,\boldsymbol{x})$  produced by the Gibbs sampler for 10,000 runs.}
\label{fig2}
\end{figure}

\begin{figure}
\includegraphics[width=80mm,height=80mm]{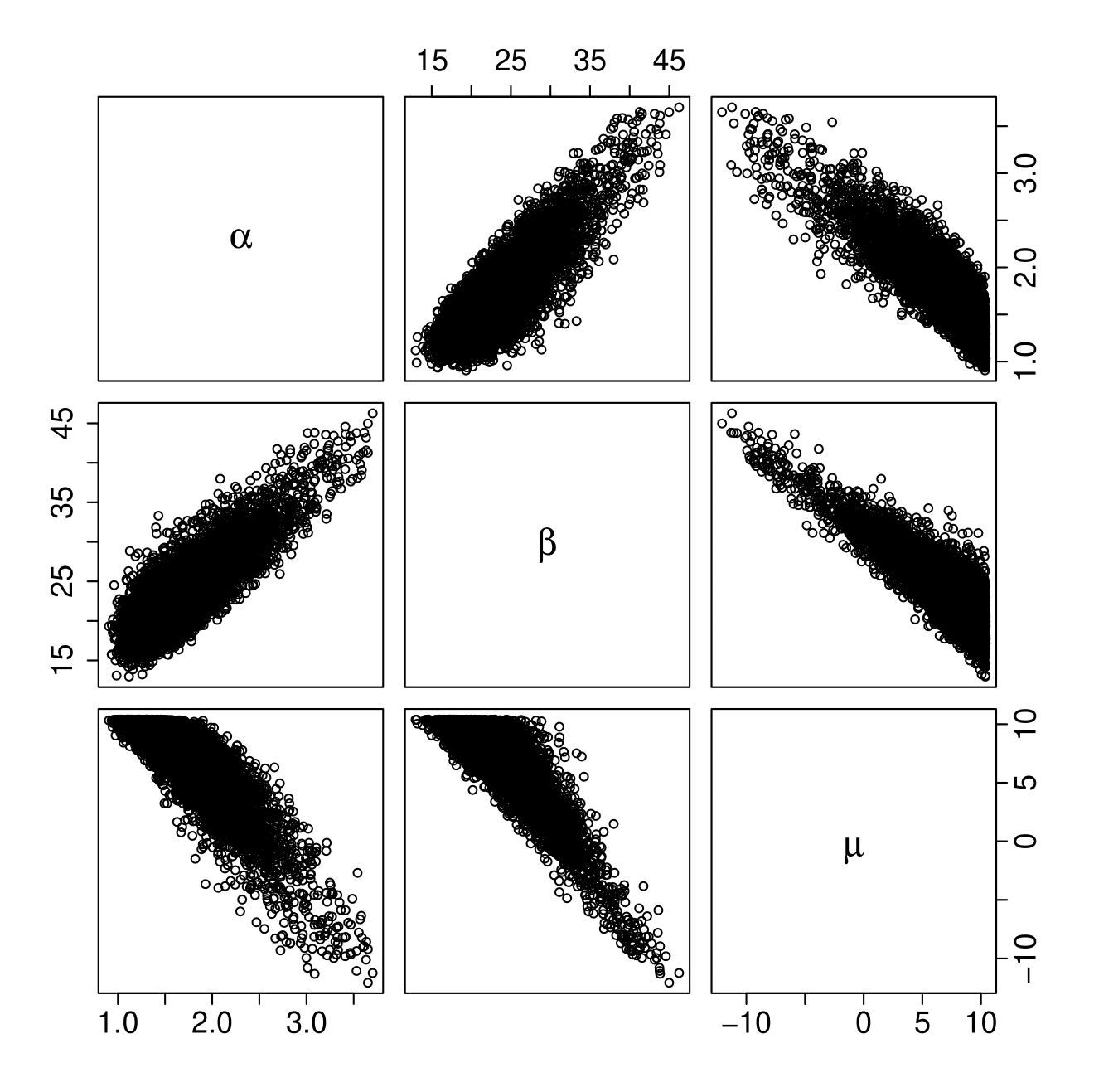}
\includegraphics[width=80mm,height=80mm]{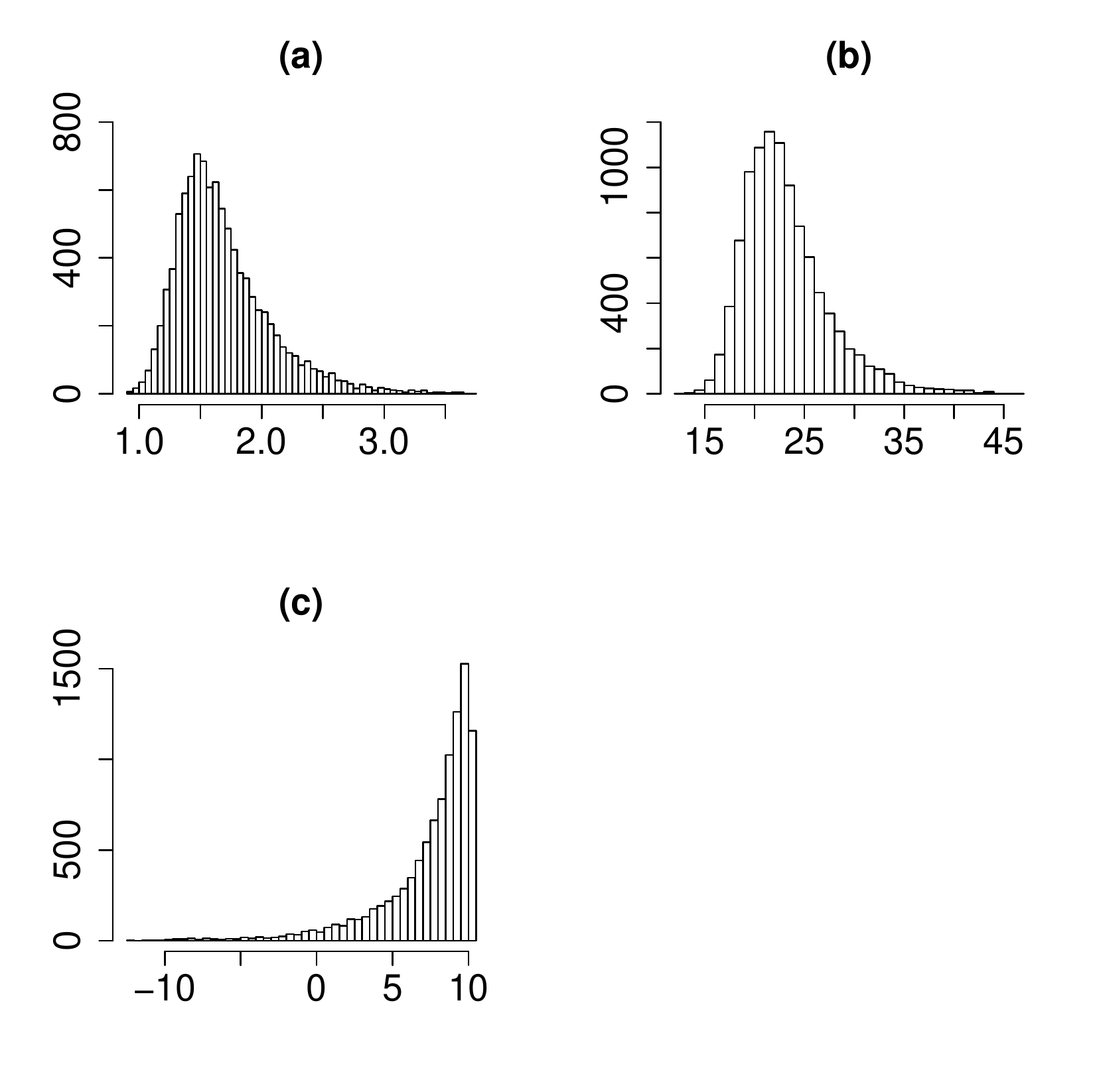}
\caption{Left-hand side: pairwise scatterplots of trimmed output of the Gibbs sampler for estimation parameters of the three-parameter Weibull distribution fitted to the DBH observations in plot 9. These outputs suggest that there is considerable dependence between all pairs. Right-hand side: histograms of the full conditionals (a) $\pi(\alpha|\beta,\mu,\boldsymbol{x})$, (b) $\pi(\beta|\alpha,\mu,\boldsymbol{x})$, and (c) $\pi(\mu|\alpha,\beta,\boldsymbol{x})$ produced by the Gibbs sampler for 10,000 runs.}
\label{fig3}
\end{figure}

\begin{figure}
\includegraphics[width=80mm,height=80mm]{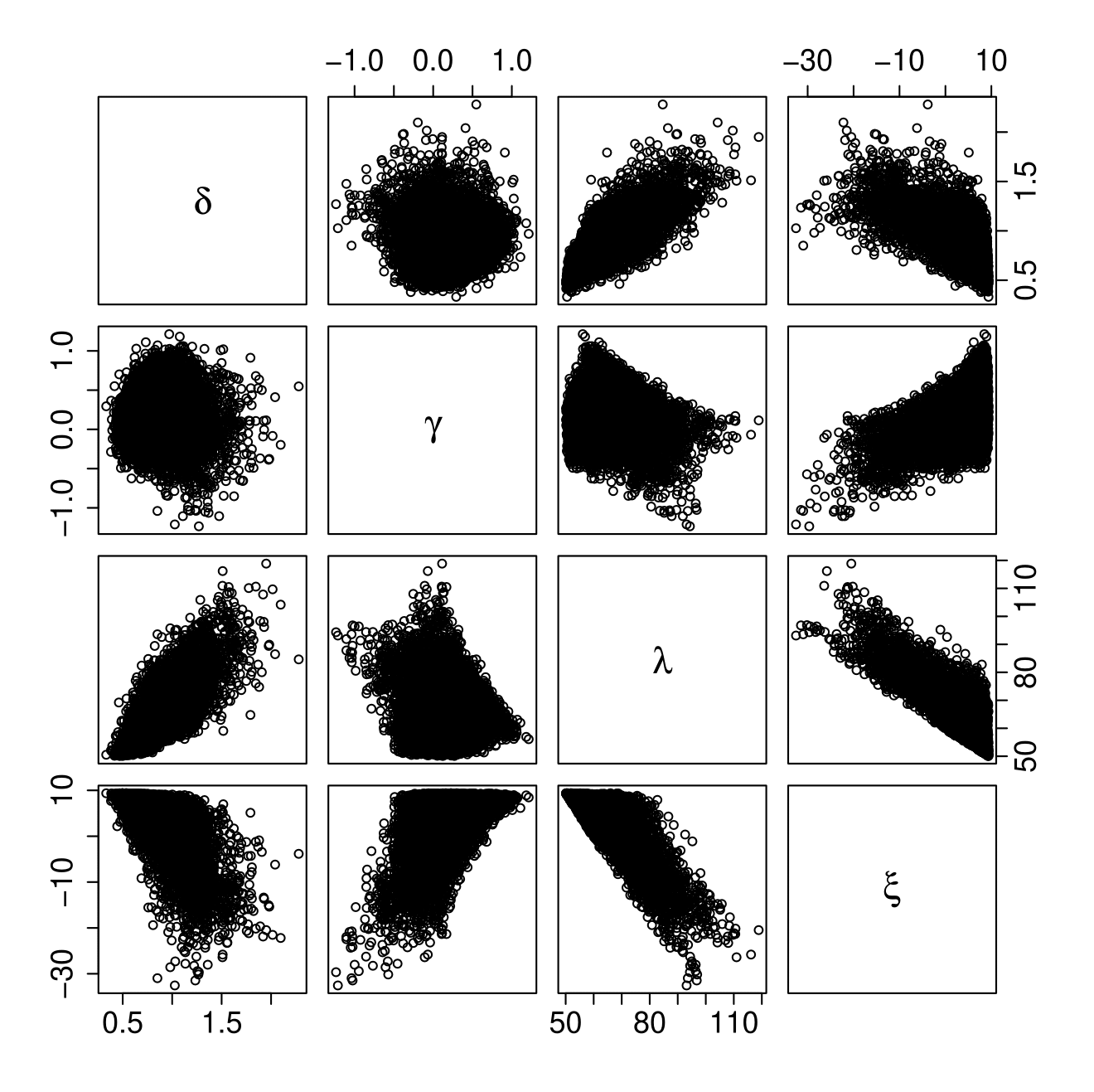}
\includegraphics[width=80mm,height=80mm]{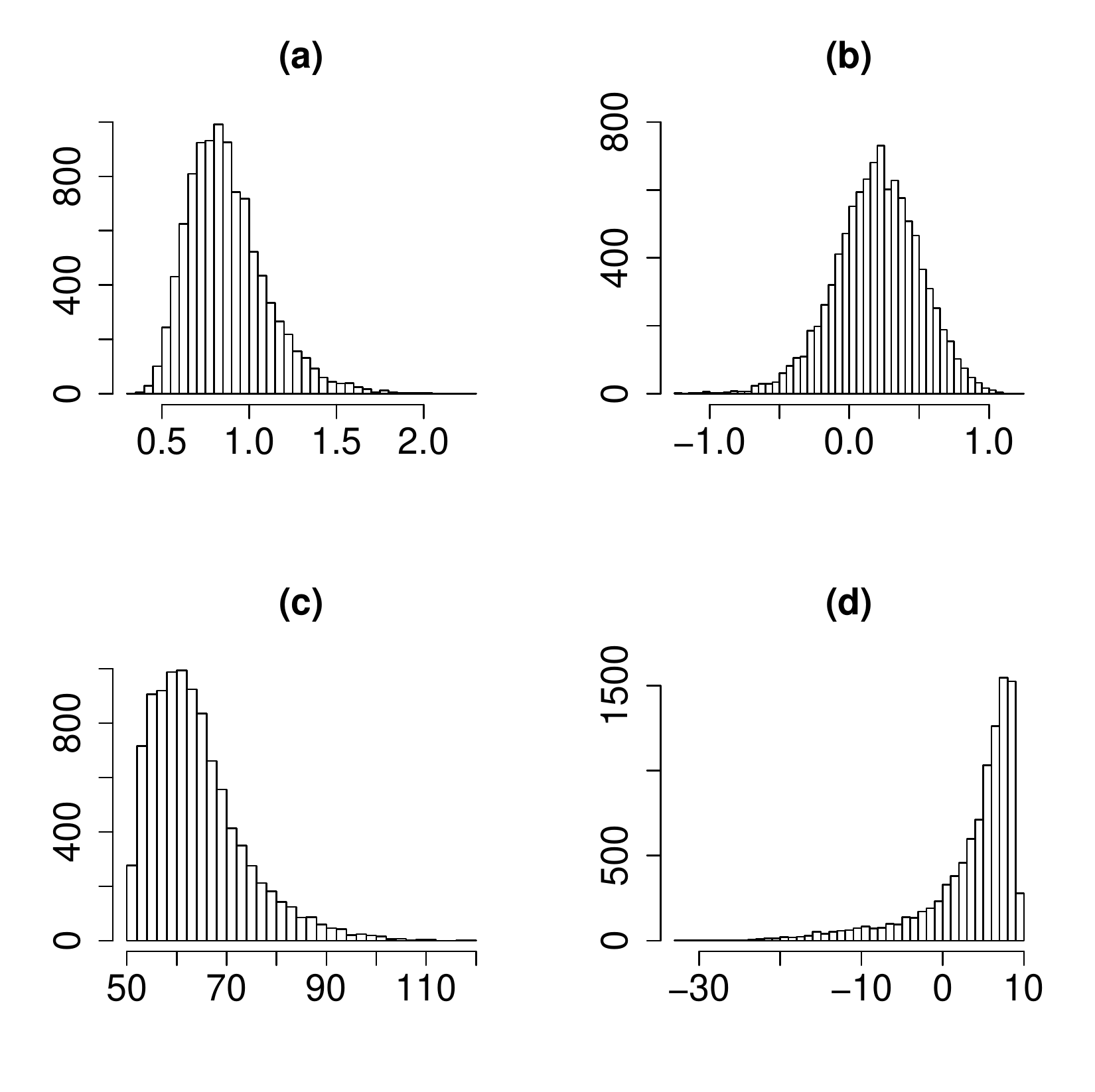}
\caption{Left-hand side: pairwise scatterplots of trimmed output of the Gibbs sampler for estimation parameters of the JSB distribution fitted to the DBH observations in plot 44. These outputs suggest that there is little dependence between between pairs $(\delta,\lambda)$ and $(\lambda,\xi)$. Right-hand side: histograms of the full conditionals (a) $\pi(\delta|\gamma,\lambda,\xi,\boldsymbol{x})$, (b) $\pi(\gamma|\delta,\lambda,\xi,\boldsymbol{x})$, (c) $\pi(\lambda|\delta,\gamma,\xi,\boldsymbol{x})$, and (d) $\pi(\xi|\delta,\gamma,\lambda,\boldsymbol{x})$  produced by the Gibbs sampler for 10,000 runs.}
\label{fig4}
\end{figure}

\begin{figure}
\includegraphics[width=80mm,height=80mm]{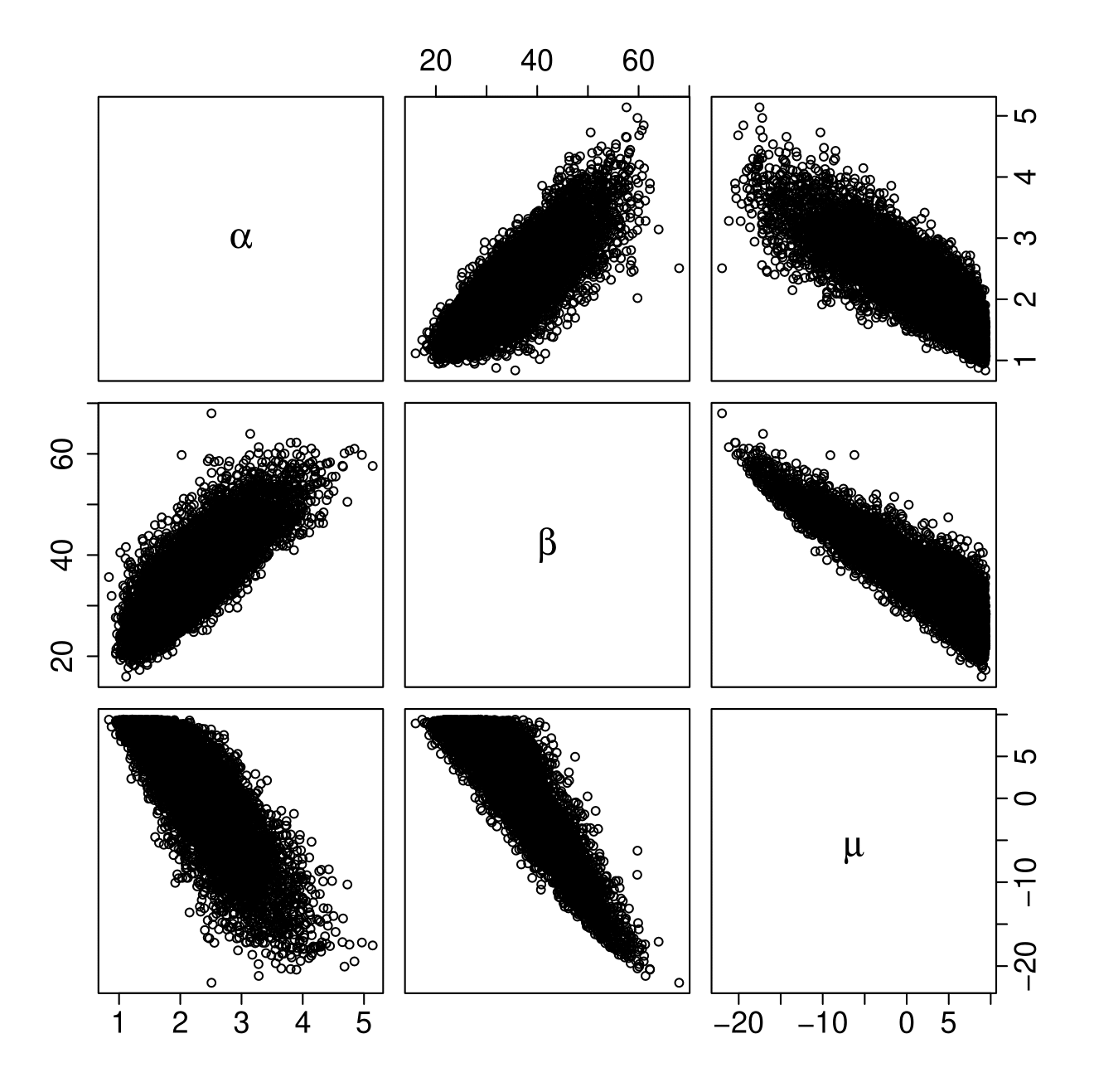}
\includegraphics[width=80mm,height=80mm]{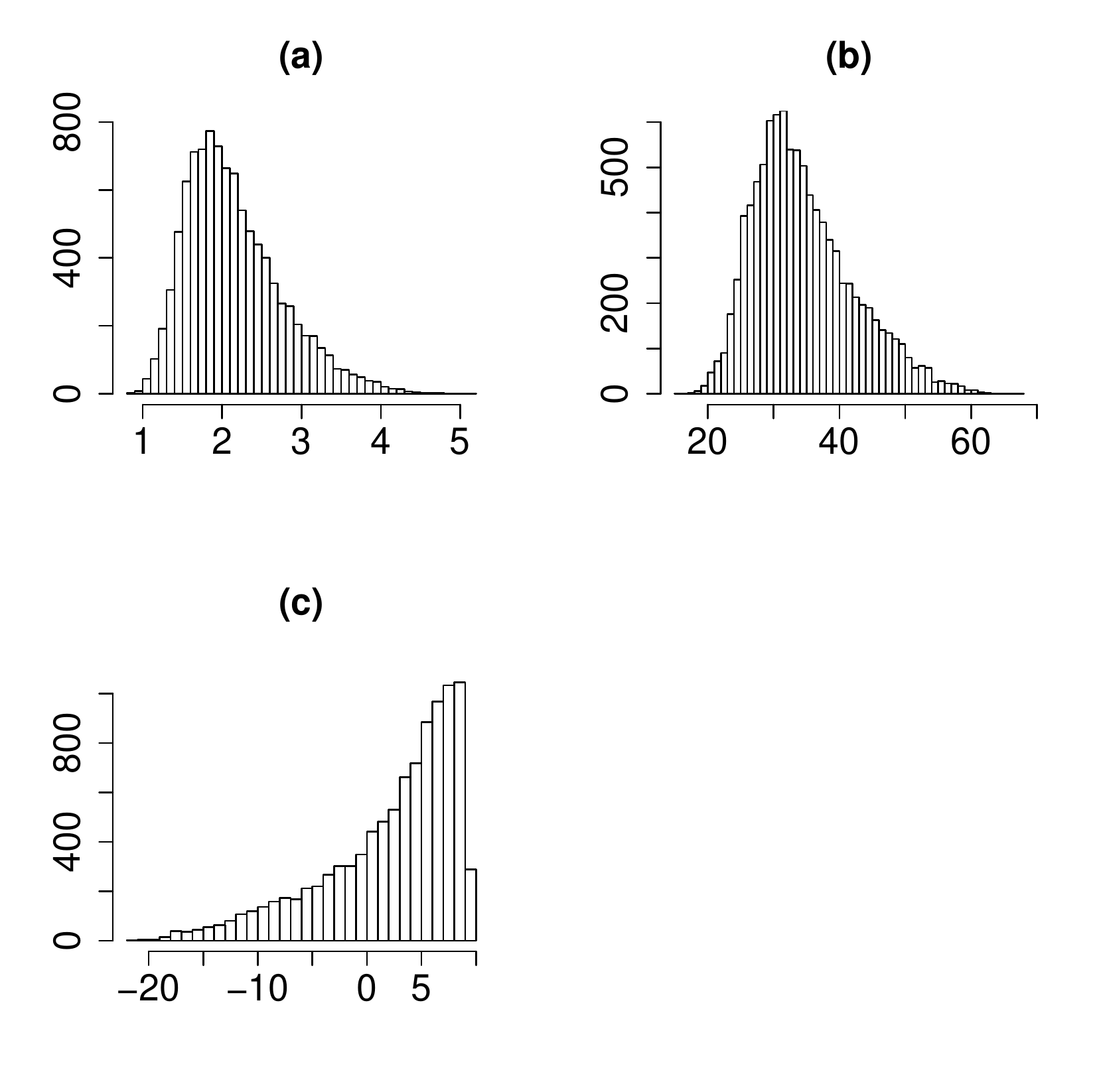}
\caption{Left-hand side: pairwise scatterplots of trimmed output of the Gibbs sampler for estimation parameters of the three-parameter Weibull distribution fitted to the DBH observations in plot 44. These outputs suggest that there is considerable dependence between all pairs. Right-hand side: histograms of the full conditionals (a) $\pi(\alpha|\beta,\mu,\boldsymbol{x})$, (b) $\pi(\beta|\alpha,\mu,\boldsymbol{x})$, and (c) $\pi(\mu|\alpha,\beta,\boldsymbol{x})$ produced by the Gibbs sampler for 10,000 runs.}
\label{fig5}
\end{figure}

\begin{figure}
\includegraphics[width=80mm,height=80mm]{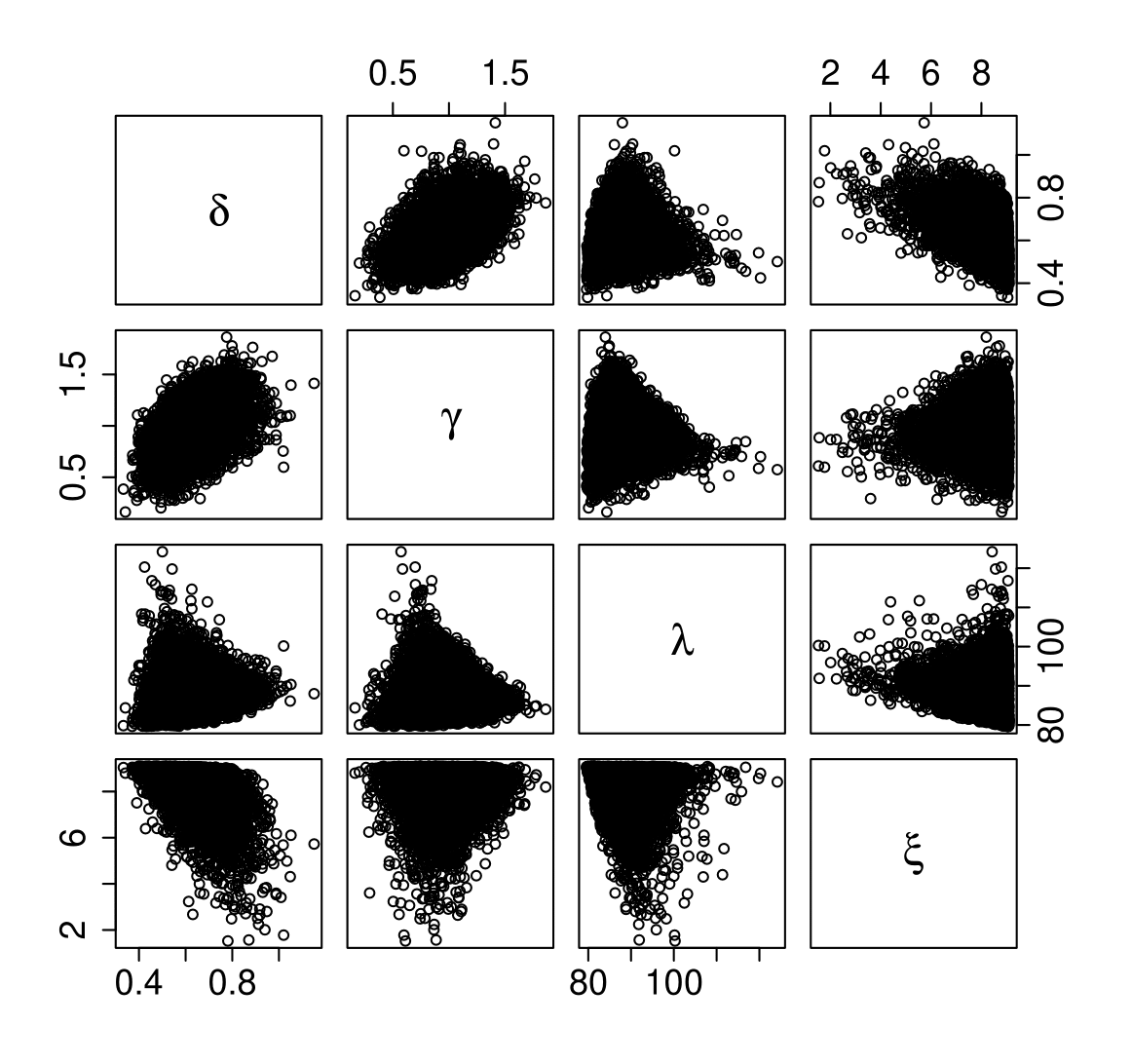}
\includegraphics[width=80mm,height=80mm]{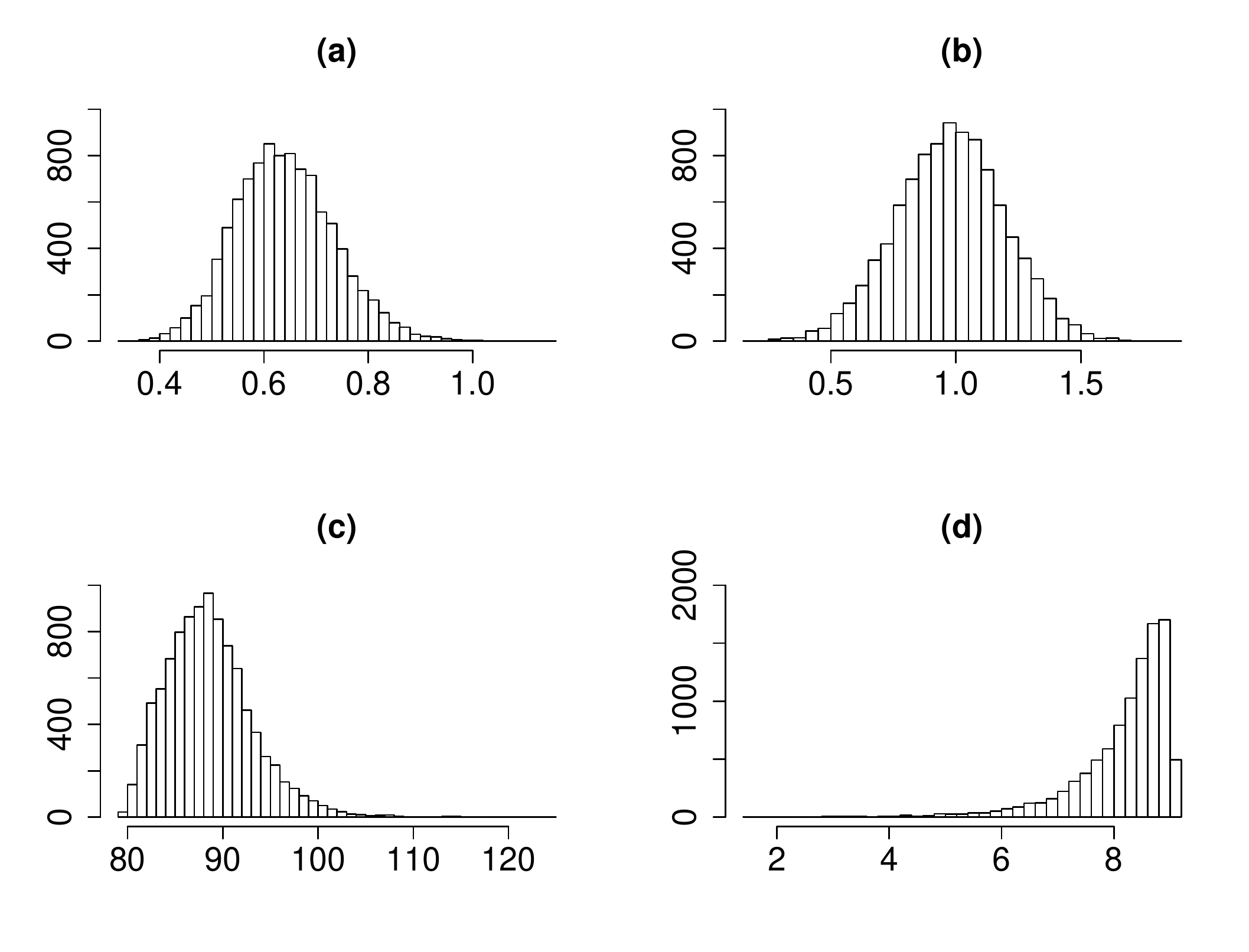}
\caption{Left-hand side: pairwise scatterplots of trimmed output of the Gibbs sampler for estimation parameters of the JSB distribution fitted to the DBH observations in plot 73. These outputs suggest that there is little dependence between $\delta$ and $\gamma$. Right-hand side: histograms of the full conditionals (a) $\pi(\delta|\gamma,\lambda,\xi,\boldsymbol{x})$, (b) $\pi(\gamma|\delta,\lambda,\xi,\boldsymbol{x})$, (c) $\pi(\lambda|\delta,\gamma,\xi,\boldsymbol{x})$, and (d) $\pi(\xi|\delta,\gamma,\lambda,\boldsymbol{x})$  produced by the Gibbs sampler for 10,000 runs.}
\label{fig6}
\end{figure}

\begin{figure}
\includegraphics[width=80mm,height=80mm]{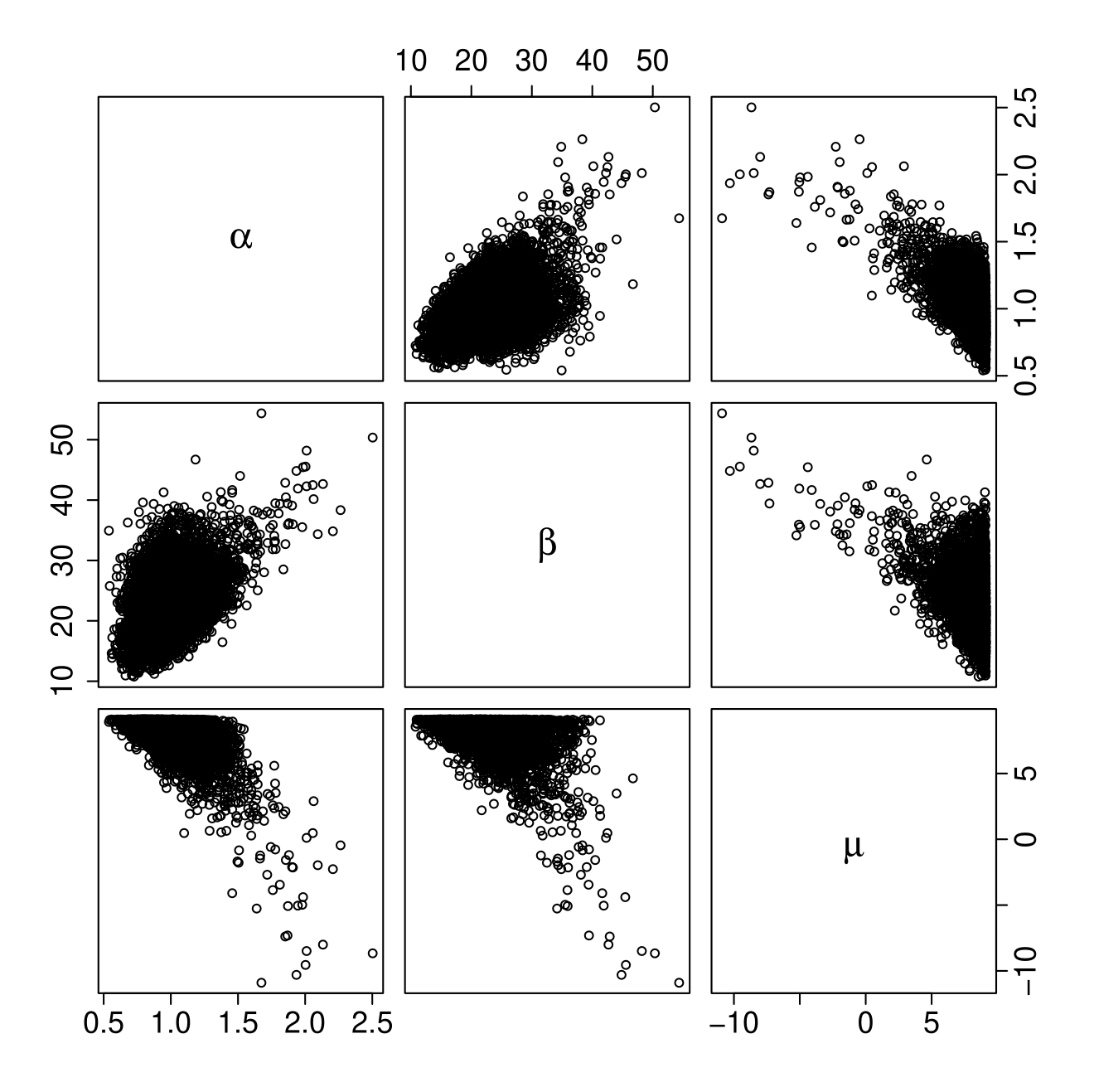}
\includegraphics[width=80mm,height=80mm]{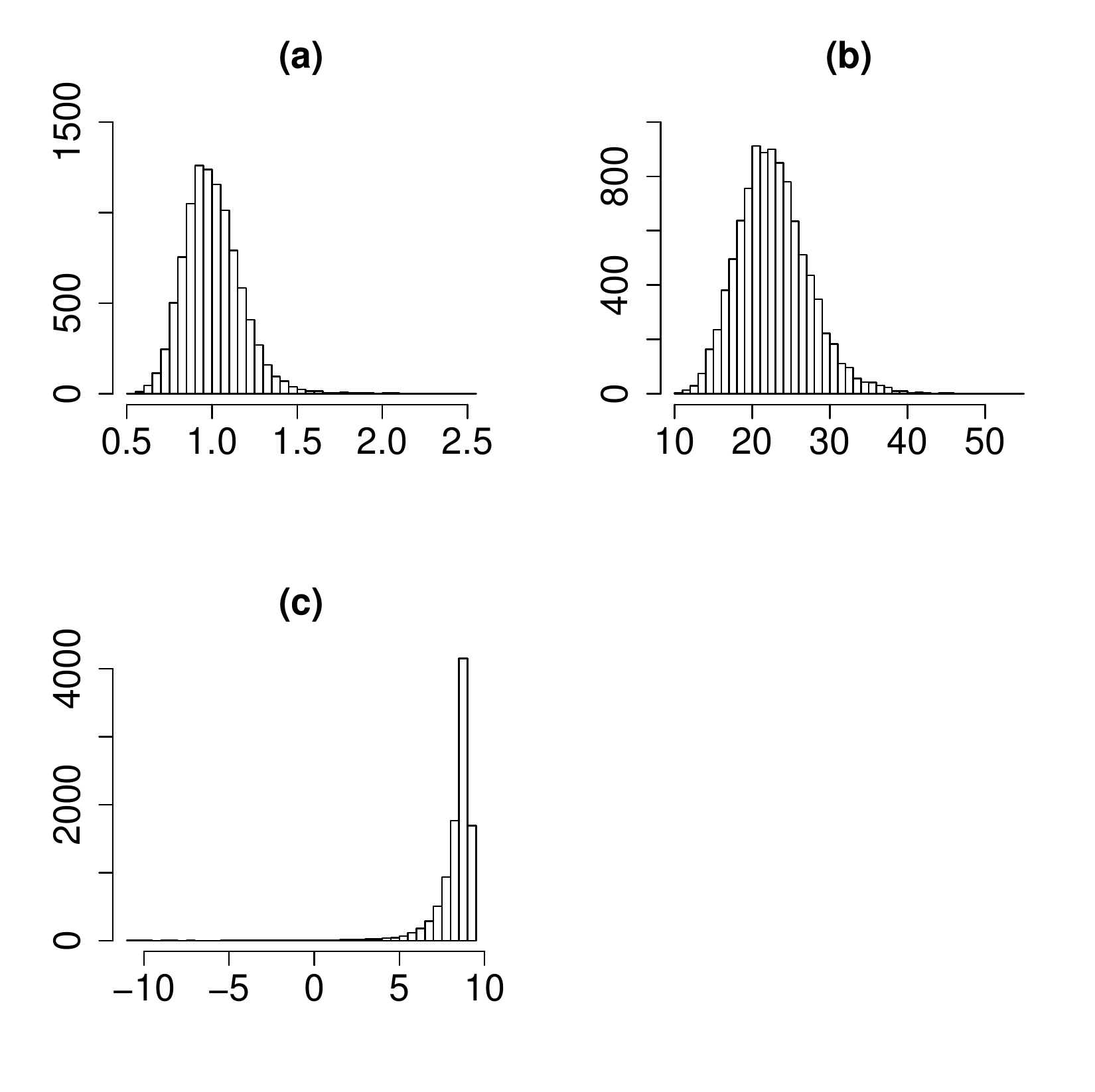}
\caption{Left-hand side: pairwise scatterplots of trimmed output of the Gibbs sampler for estimation parameters of the three-parameter Weibull distribution fitted to the DBH observations in plot 73. These outputs suggest that there is little dependence between $\alpha$ and $\mu$. Right-hand side: histograms of the full conditionals (a) $\pi(\alpha|\beta,\mu,\boldsymbol{x})$, (b) $\pi(\beta|\alpha,\mu,\boldsymbol{x})$, and (c) $\pi(\mu|\alpha,\beta,\boldsymbol{x})$ produced by the Gibbs sampler for 10,000 runs.}
\label{fig7}
\end{figure}

The Bayesian approach were applied to the DBH data addressed in the previous subsection. We compared the performance of the JBS and three-parameter Weibull distributions for modelling DBH data when the parameters of both models were estimated using the Bayesian approach. Figure \ref{fig2} and Figure \ref{fig3} display histograms of the samples drawn from the full conditionals and the pairwise scatterplots of the sampler output for the JSB and three-parameter Weibull distributions, respectively, when these distributions fitted to DBH data of the plot 9. Accordingly, Figure \ref{fig4} and Figure \ref{fig5} show histograms of the samples drawn from the full conditionals and pairwise scatterplots of the sampler output for the JSB and three-parameter Weibull distributions, respectively, when these distributions fitted to the DBH data in the plot 81. Finally, the graphical visualizations when the JSB and three-parameter Weibull distributions fitted to the DBH data in the plot 73 are shown in Figure \ref{fig6} and Figure \ref{fig7}, respectively. We assumed that the sampler's convergence has been attained before 5,000 iterations in all three plots for both the JSB and three-parameter Weibull distributions. Therefore, we removed the first 5000 samples from the sampler output when the sampler were repeated for 10,000 times. Based on the average of the final 5,000 samples, we obtained the Bayesian estimators of the models parameters. The estimated parameters are given in Table \ref{tab4}. The goodness-of-fit statistics including Anderson-Darling (AD), Cram\' {e}r–von Mises (CM), Kolmogorov-Smirnov (KS), and log-likelihood (LL) statistics were used as criterion for choosing the better model. These criteria are defined as
\begin{align*}
AD&=-n-\sum_{i=1}^{n}\frac{2i-1}{n}\biggl[\log G\bigl(x_{(i)}\big|\hat{\Theta}\bigr)+\log\Bigl(1-G\bigl(x_{(n+i-1)}\big|\hat{\Theta}\bigr)\Bigr)\biggr],\nonumber\\
CM&=\frac{1}{12n}+\sum_{i=1}^{n}\biggl[G\bigl(x_{(i)}\big|\hat{\Theta}\bigr)-\frac{2i-1}{2n}\biggr]^2,\nonumber\\
KS&=\underset{x_i \in {\boldsymbol{x}}}{\operatorname{sup}}  \Big| G\bigl(x_i\big|\hat{\Theta}\bigr)-G_{n}(x_i)\Big|,\\
LL&=\sum_{i=1}^{n} \log g\bigl(x_{i}\big|\hat{\Theta}\bigr),
\end{align*}
where $G_{n}(.)$ denotes the empirical distribution function, $x_{(i)}$ is th $i$-th order statistic in the random sample of size $n$, and $\hat{\Theta}$ is the estimated vector parameter. The computed goodness-of-fit statistics are given in Table \ref{tab5}. 
For analyzing the DBH data given in Table \ref{tab2}, we used the R package \verb|ForestFit| \citep{teimouri2020forestfit} developed
for R \citep{2018} and uploaded to CRAN (Comprehensive R Archive Network) at \verb+https://cran.r-project.org/web/packages/Fores+\\
\verb+tFit/index.html+.
\section{Results}\label{sec4}
\begin{figure}
\centering
\includegraphics[width=52mm,height=50mm]{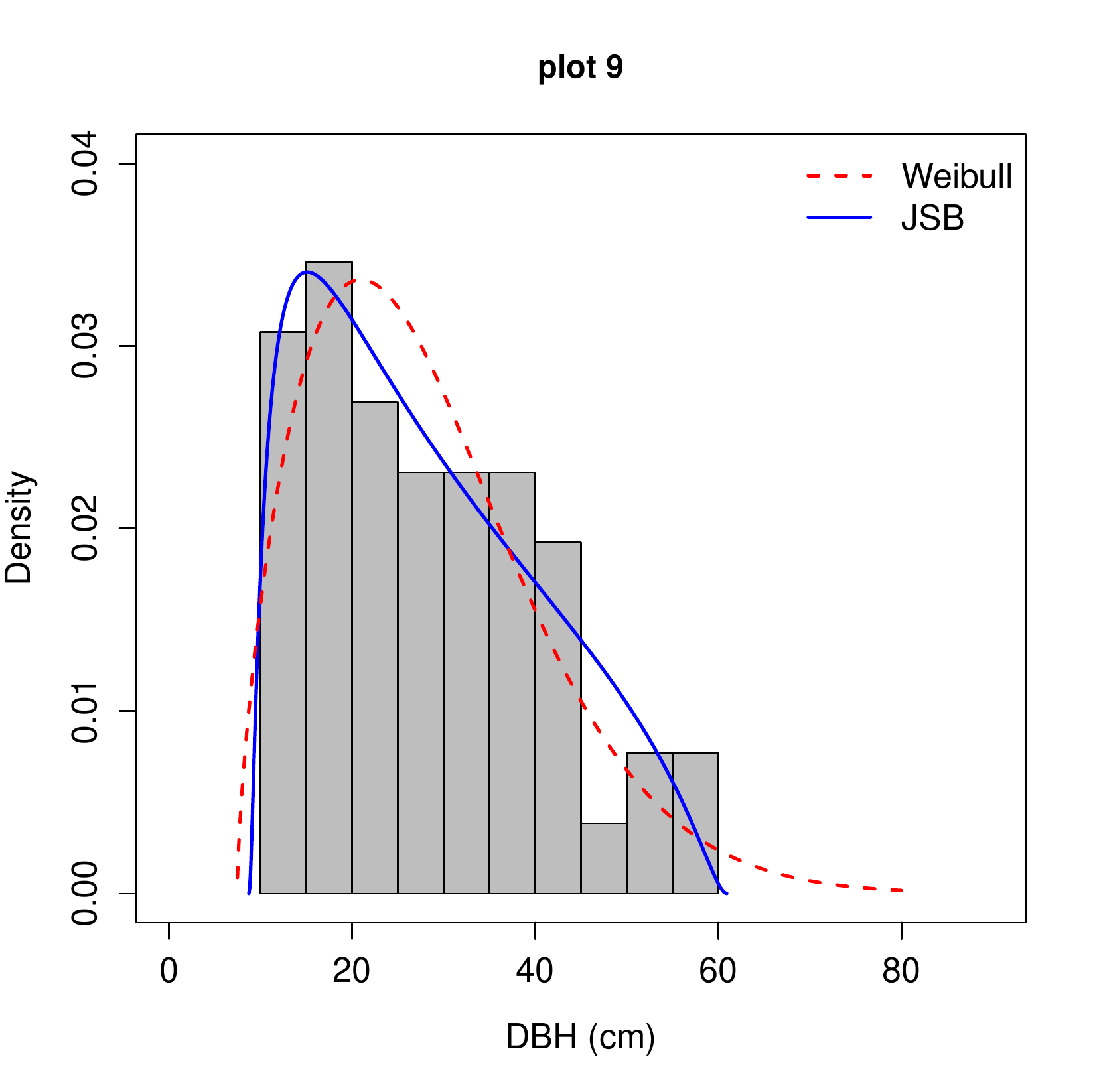}
\includegraphics[width=52mm,height=50mm]{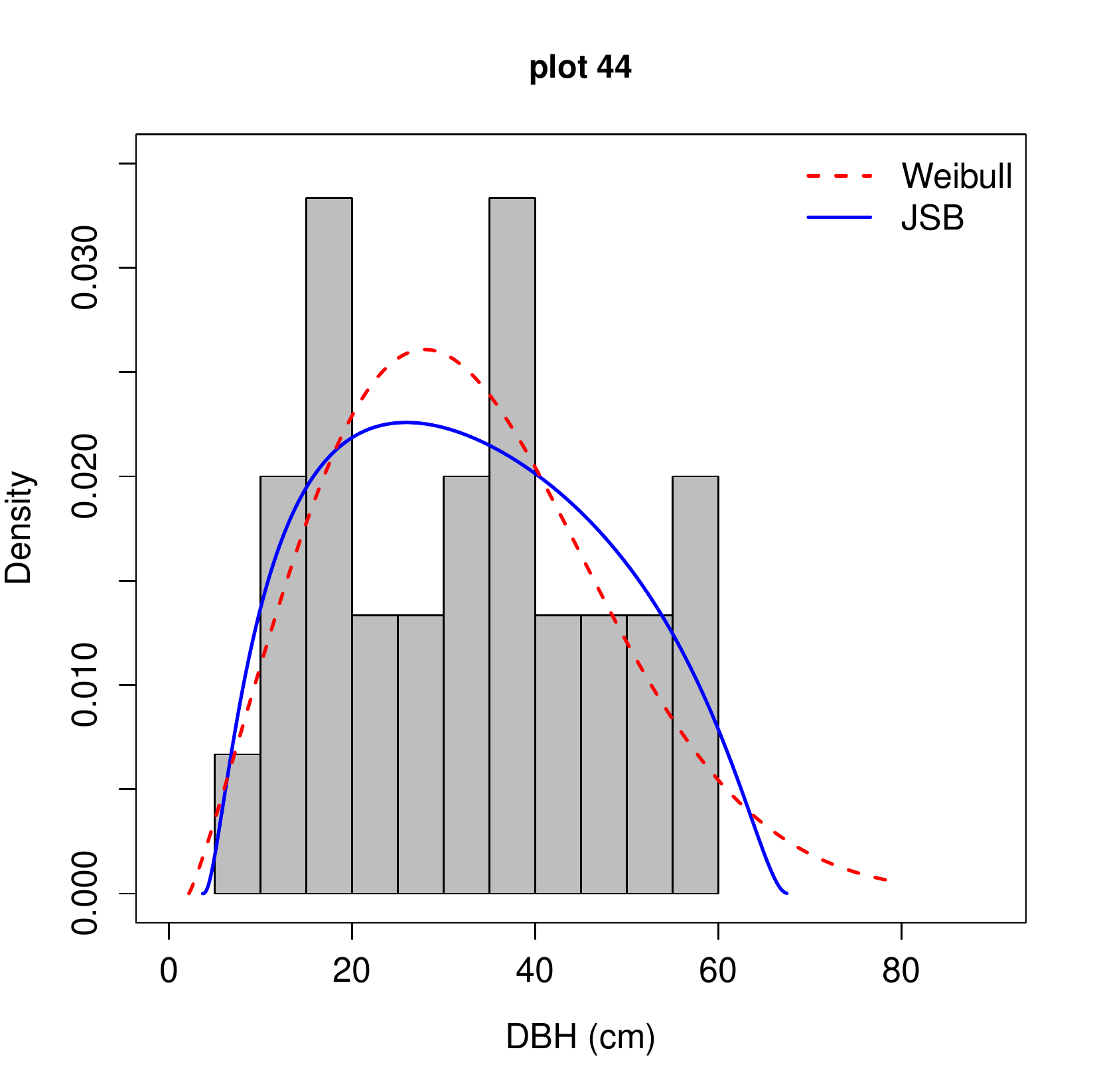}
\includegraphics[width=52mm,height=50mm]{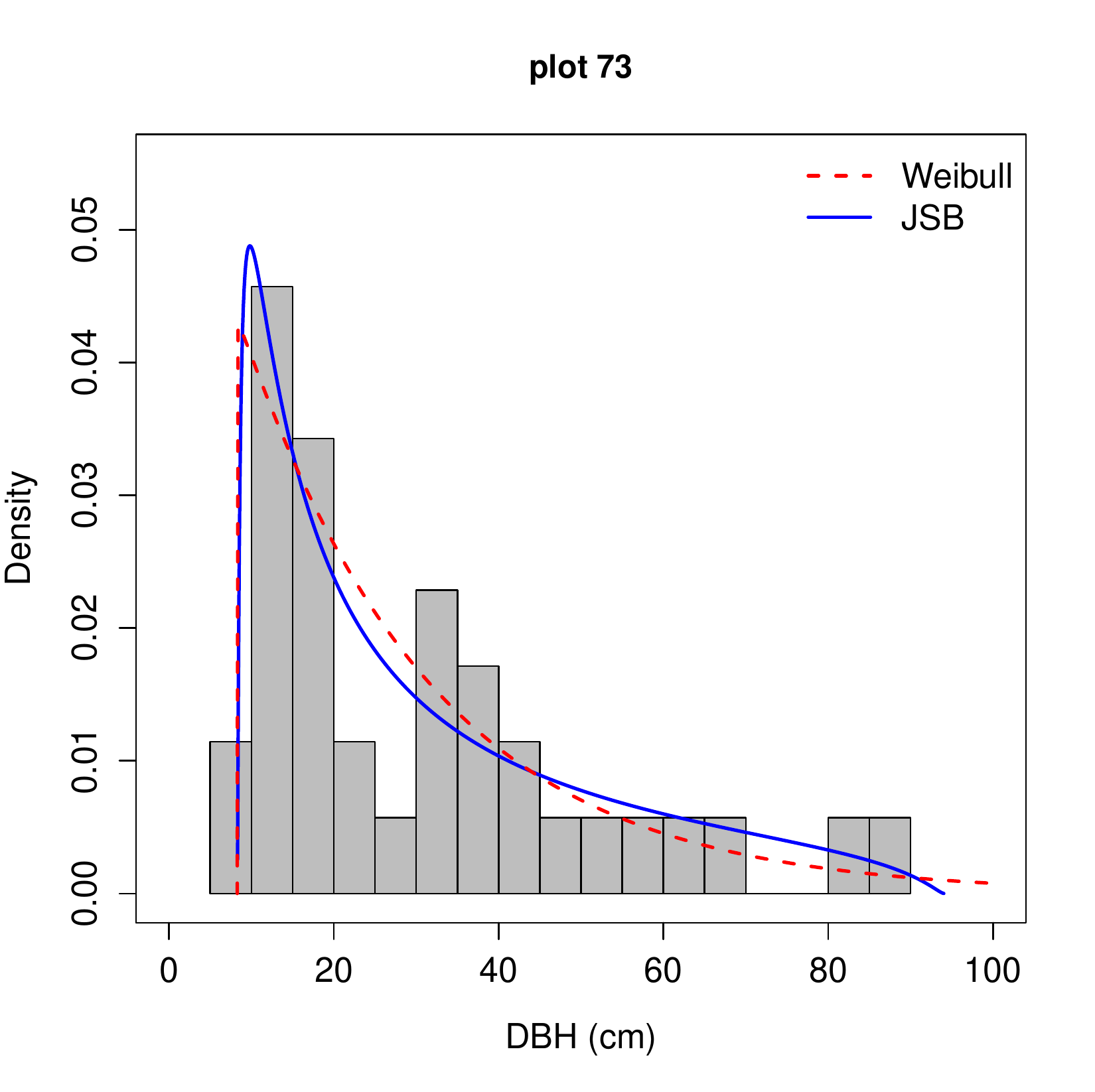}
\caption{Histograms of DBH data in plots 9, 44, and 73. Superimposed in each subfigure are estimated probability density functions of the JSB (blue solid line) and Weibull (red dashed line) distributions.}
\label{fig8}
\end{figure}

The DBH values ranged from 9.1 to 88.6 cm and the mean of DBH ranged between 24.27 and 33.35 cm in DBH (Table \ref{tab2}). The DBH distributions were usually continuous with peaks at the lower, middle, and higher bins (Fig. \ref{fig8}). Mixed forests with ponderosa pine and scattered western junipers are usually a mosaic composed of different forest patches. The analysis of three randomly
selected plots represented the following results.
\par
Plot 9 was characterized by DBH distribution that had trees in the size between 10.4 and 55.9 cm in DBH (Fig. \ref{tab2}). 
The understory cohort was characterized by a distribution with an understory cohort that has one peak around 15 cm in DBH, and truncated at 60 cm in DBH (Figure \ref{fig8}). Overall, the JSB model was the superior model than the Weibull model for DBH distribution (Table \ref{tab4}). 
The DBH distribution was best captured better by JSB distribution than the Weibull. The JSB was the superior model in the sense of AD, CM, KS, and LL measures (Table \ref{tab4}). The shapes of the fitted distributions to the DBH data of plot 9 were different in the right tail (Fig. \ref{fig8}). In fact, the JSB was characterized the tail of the DBH distribution better than the Weibull distribution.
\par
Plot 44 was characterized by DBH distribution that had trees in the size between 11.9 and 83.8 cm in DBH (Table \ref{tab2}). This plot represented a one-storied forest that was characterized by a DBH distribution with three modes around 15, 35, and 60 cm in DBH. Ignoring these modes, the understory cohort was characterized by a broadly uniform distribution with an understory cohort between 9-60 cm in DBH (Figure \ref{fig8}). Overall, the JSB model was the superior model than the Weibull model for DBH distribution (Table \ref{tab4}). The shape of the fitted densities corresponding to the JSB and Weibull distributions have differences in the middle and the right tail (Figure \ref{fig8}). This is largely because the JSB has two threshold parameters that makes its pdf bounded from the left and right.
\par
Plot 73 was characterized by DBH distribution that had trees in the size between 9.1 and 88.6 cm in DBH (Table \ref{tab2}). This plot was characterized by a generally negative exponential or reverse-\emph{J} DBH distribution. Overall, this type of DBH distribution occurs in uneven-aged, complex structure, and old-growth forests where the number of trees declines sharply with increasing tree size (Figure \ref{fig8}). There are two peaks around 15 and 35 cm in DBH. In contrast to the plot 9 and 44, the DBH distribution of plot 73 was much wider with a long and skewed to the right tail. The JSB model was characterized the DBH distribution better than the Weibull model (Table \ref{tab4}). Evidently, the JSB was the superior model than the Weibull model for DBH observations at the first peak, between two peaks, and right tail (Figure \ref{fig8}). In all three plots, the JSB outperformed the Weibull model in terms of all AD, CM, KS, and LL measures (Table \ref{tab4}).
\section{Discussion}\label{sec5}
For implementing the Bayesian paradigm, we carried out a study for choosing the initial values. In the case of the JSB distribution, it is known that the first order statistic, i.e., $x_{(1)}$ is a sufficient statistic for $\xi$. So, choosing $\xi^{(0)}=x_{(1)}-1/n$ in which $n$ is the sample size would be quite reasonable as the initial value for $\xi$. In the same fashion as for $\xi$, a good initial value for $\lambda$, since $\xi<x<\xi+\lambda$, is given by $\lambda^{(0)}=x_{(n)}-x_{(1)}+2/n$ in which $x_{(n)}$ is the  maximum value of DBH in the sample, i.e., $x_{(n)}$. Here, the constants $1/n$ and $2/n$ for $\xi^{(0)}$ and $\lambda^{(0)}$ has been used to avoid the possible singularity problems. A suitable initial value for $\gamma$ is obtained by considering the relation $\gamma=\delta \log \bigl(1/{\boldsymbol{y}}_{0.5}-1\bigr)$ where ${\boldsymbol{y}}_{0.5}$ is median of the transformation ${\boldsymbol{y}}=({\boldsymbol{x}}-\xi)/\lambda$ with ${\boldsymbol{x}}=\bigl(x_1,\dots,x_n\bigr)^T$ \citep{ozccelik2016modeling}. Therefore, $\gamma^{(0)}=\delta^{(0)}\log \bigl(1/{\boldsymbol{y}}_{0.5}-1\bigr)$ where ${\boldsymbol{y}}_{0.5}$ is median of the transformation ${\boldsymbol{y}}=\bigl({\boldsymbol{x}}-\xi^{(0)}\bigr)/\lambda^{(0)}$. It should be noted that we took $\delta^{(0)}=1$ as the initial value for $\delta$. 
\par
\begin{figure}[!h]
\centering{
\includegraphics[width=110mm,height=110mm]{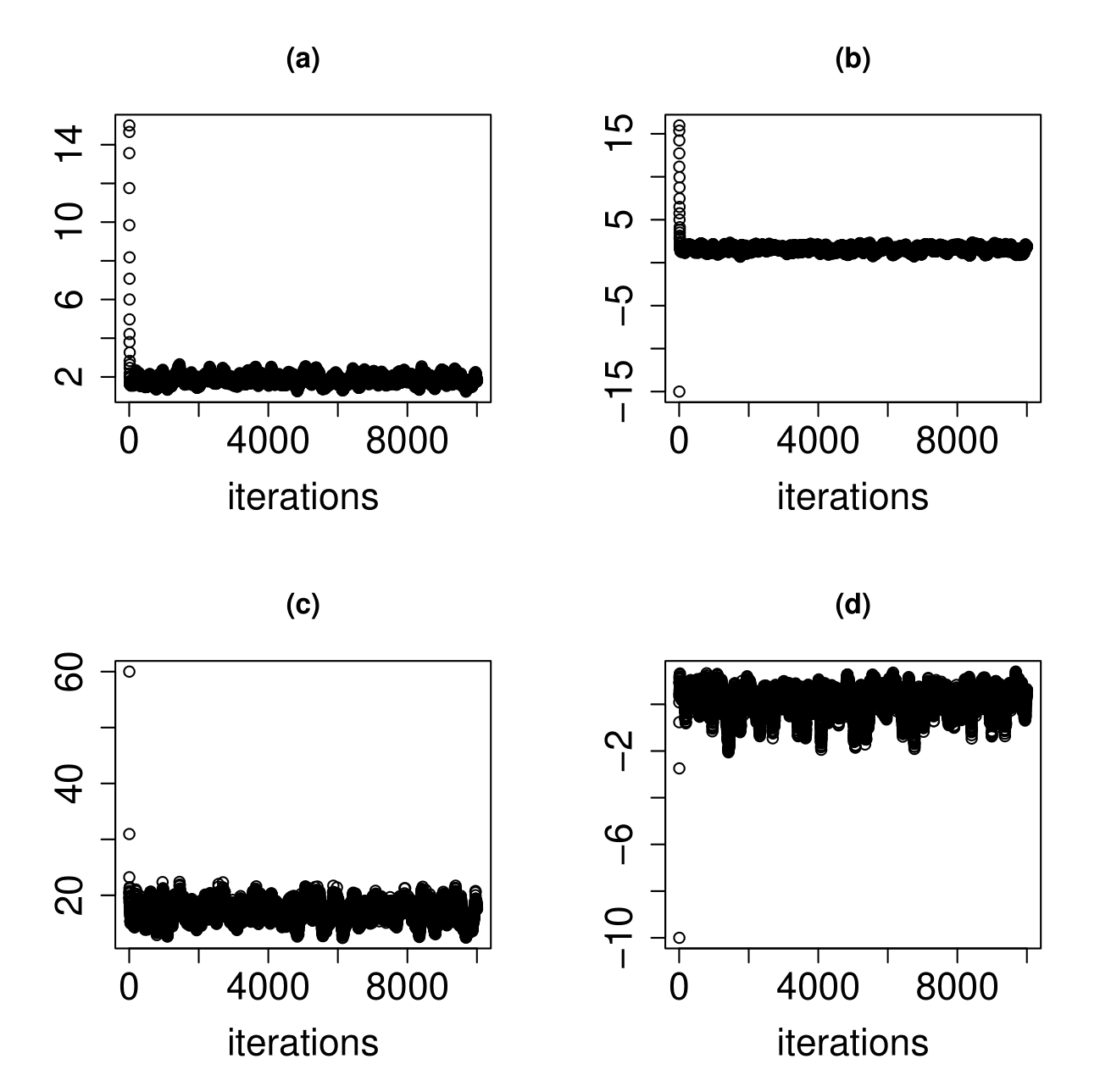}}
\caption{Output of (a) $\delta$, (b) $\gamma$, (c) $\lambda$, and (d) $\xi$ from Gibbs sampler for the robust analysis. Each subfigure shows the motion of the Gibbs sampler across  10,000 iterations.}
\label{fig9}
\end{figure}

 Additionally, our study revealed that  the Bayesian paradigm presented in this work is robust with respect to $\delta^{(0)}$, so that it can be started well away from the true value of $\delta$. In the case of the three-parameter Weibull distribution, similar to the JSB distribution, we used $\xi^{(0)}=x_{(1)}-1/n$ as the initial value for $\mu$. The initial values of the shape and scale parameters obtained by using the method of moments \citep{norman1994continuous}. We also performed a simulation study to check the robustness of the Bayesian paradigm with respect to the initial values for estimating the parameters of the JSB distribution. For this purpose, we confine ourselves to the case in which we have simulated 300 samples each of size 100 from JSB distribution with parameter vector $\Theta=(2,2,20,0)^T$, i.e., $\delta=2$, $\gamma=2$, $\lambda=20$, and $\xi=0$. The results of simulation are given in Table \ref{tab3}. 
We note that in each of 200 runs, the initial values were not chosen by the method suggested above. Instead, the initial values were generated randomly from uniform distribution. We used this scenario in order to check the robustness of the Gibbs sampler. The initial values for $\delta$, $\gamma$, $\lambda$ ,and $\xi$ were generated from uniform distribution (0.1,15), (-15,15), (20.1,60), and (-10,10), respectively. For example, the general motion of the Gibbs sampler has been shown in Figure \ref{fig9}, when the initial values were chosen as $\delta^{(0)}=15$, $\gamma^{(0)}=-15$, $\lambda^{(0)}=60$, and $\xi^{(0)}=-10$ to show the robustness of the Bayesian paradigm.
\section{Conclusion}\label{sec6}
We have derived the Bayesian estimators for the four-parameter Johnson's SB (JSB) distribution. The maximum likelihood (ML) approach is the most commonly used method for estimating the model parameters, but it has been shown using simulation study that percentage of failed attempts to reach the convergence through this method is on the average 32\%. So, we suggest to use the Bayesian paradigm to estimate the parameters of the JSB distribution. We have shown that the proposed Bayesian approach works efficiently and is robust with respect to the initial values. Furthermore, we have considered the Bayesian estimators for parameters of the three-parameter Weibull distribution that proposed by Green et al. (1994). Our algorithm for sampling from full conditional of the location parameter is faster to that of Green et al. (1994). The We have fitted both of the JSB and three-parameter Weibull distributions to the diameters at breast height (DBH) obtained from 3 plots out of 107 plots of size 0.08 ha established in mixed-age ponderosa pine ({\it{Pinus ponderosa}} Dougl. ex Laws.) forests with scattered western junipers located in the Malheur National Forest on the south end of the Blue Mountains near Burns, Oregon, USA. The estimation results indicated that the JBS model outperformed the three-parameter Weibull distribution and so characterized more accurately the DBH distribution. As a possible future work, we are interested in estimating the parameters of the bivariate JSB distribution using the Bayesian method. The users can access the \verb|R| package \verb|ForestFit| that is available at address \verb|https://cran.r-project.org/web/packages/ForestFit/index.html|. 
\appendix
\section{}\label{apa}
We have
\begin{align*}
\pi(\delta| \gamma, \lambda, \xi, \boldsymbol{x}) &\propto \frac{\delta^n\lambda^n}{(2\pi)^{\frac{n}{2}}\Pi_{i=1}^{n}(x_i-\xi)(\lambda+\xi-x_i)}
\exp\left\{-\frac{1}{2}\sum_{i=1}^{n}\Bigg[\gamma+\delta \log\biggl(\frac{x_i-\xi}{\lambda+\xi-x_i}\biggr)\Bigg]^2\right\} \nonumber\\
&\propto \delta^n \exp\left\{-\frac{k_2}{2}\biggl[\delta+\frac{\gamma k_1}{k_2}\biggr]^2\right\},
\end{align*}
where 
\begin{align}\label{k1}
k_1=\sum_{i=1}^{n}\log\Bigl(\frac{x_i-\xi}{\lambda+\xi-x_i}\Bigr),
\end{align}
and
\begin{align}\label{k2}
k_2=\sum_{i=1}^{n}\bigg[\log\Bigl(\frac{x_i-\xi}{\lambda+\xi-x_i}\Bigr)\biggr]^2. 
\end{align}
The full conditional pdf of $\delta$ is given by 
\begin{align*}
\pi(\delta| \gamma, \lambda, \xi, \boldsymbol{x}) =\text{C} \delta^n \exp\left\{-\frac{k_2}{2}\biggl[\delta+\frac{\gamma k_1}{k_2}\biggr]^2\right\},
\end{align*}
where $\text{C}$ is a normalizing constant independent of $\delta$. The First and second derivatives of $\log \pi(\delta| \gamma, \lambda, \xi, \boldsymbol{x}) $ with respect to $\delta$ are
\begin{align*}
\frac{\partial \pi(\delta| \gamma, \lambda, \xi, \boldsymbol{x})}{\partial \delta}=\frac{n}{\delta}-k_2\biggl[\delta+\frac{\gamma k_1}{k_2}\biggr],
\end{align*}
and
\begin{align}\label{seconddelta}
\frac{\partial^2 \pi(\delta| \gamma, \lambda, \xi, \boldsymbol{x})}{\partial \delta^2}=-\frac{n}{\delta^2}-k_2.
\end{align}
Since $k_2>0$, the right-hand side of (\ref{seconddelta}) is always negative and so $\pi(\delta| \gamma, \lambda, \xi, \boldsymbol{x})$ is log-concave. Assume that we are currently at $t$-th iteration of the sampler, sampling from $\pi\bigl(\delta \big| \gamma^{(t)}, \lambda^{(t)}, \xi^{(t)}, \boldsymbol{x}\bigr)$ is carried out through the ARS algorithm. Here, $\gamma^{(t)}$, $\lambda^{(t)}$, and $\xi^{(t)}$ denote the generated values, respectively, from $\gamma$, $\lambda$, and $\xi$, at $t$-th iteration.
\section{}\label{apb}
We have
\begin{align*}
\pi(\gamma| \delta, \lambda, \xi, \boldsymbol{x}) \propto \exp\left\{-\frac{1}{2}\sum_{i=1}^{n}\Bigg[\gamma+\delta \log\biggl(\frac{x_i-\xi}{\lambda+\xi-x_i}\biggr)\Bigg]^2\right\} \propto \exp\left\{-\frac{n}{2}\biggl[\gamma+\frac{\delta k_1}{n}\biggr]^2\right\},
\end{align*}
where $k_1$ is defined in (\ref{k1}). Assume that we are currently at $t$-th iteration of the sampler, for sampling from full conditional pdf of $\gamma$, it is enough to sample from Gaussian distribution with mean
\begin{align*}
-\frac{\delta^{(t+1)}}{n}\sum_{i=1}^{n}\log\biggl(\frac{x_i-\xi^{(t)}}{\lambda^{(t)}+\xi^{(t)}-x_i}\biggr),
\end{align*}
and variance $1/n$.
\section{}\label{apc}
It  is easy to see that the full conditional of $\lambda$ is (up to proportionality)
\begin{align*}
\pi(\lambda|\delta, \gamma, \xi, \boldsymbol{x}) &\propto \Pi_{i=1}^{n}\biggl(\frac{\lambda}{\lambda+\xi-x_i}\biggr)
\exp\left\{-\frac{1}{2}\sum_{i=1}^{n}\Bigg[\gamma+\delta \log\biggl(\frac{x_i-\xi}{\lambda+\xi-x_i}\biggr)\Bigg]^2\right\} \nonumber\\
&= \lambda^n \Pi_{i=1}^{n} \Bigl(\lambda+\xi-x_i\Bigr)^{\delta \gamma-1} \exp\left\{-\frac{\delta^2}{2}\sum_{i=1}^{n}\Bigg[\log\biggl(\frac{x_i-\xi}{\lambda+\xi-x_i}\biggr)\Bigg]^2\right\},
\end{align*}
As it is seen, the structure of the full conditional pdf of $\lambda$ is complicated. Additionally, $\pi(\lambda|\delta, \gamma, \xi, \boldsymbol{x})$ is not always log-concave. So, we use the MH algorithm for sampling from $\pi(\lambda|\delta, \gamma, \xi, \boldsymbol{x})$ by choosing $\exp\Bigl\{-\bigl(\lambda-x_{(n)}+\xi\bigr)\Bigr\}$ as the proposal pdf for $\lambda>x_{(n)}-\xi$ wherein $x_{(n)}=\max\{x_1,\dots,x_n\}$. The six-step MH algorithm is given a follows.
\begin{enumerate}
\item Suppose we are currently at $t$-th iteration of the sampler. Choose the initial value as $\lambda_{(0)}=x_{(n)}-\xi^{(t)}+1/n$ and set i=1;
\item Sample $\lambda^{*}$ from proposal distribution $q\bigl(\lambda)$ with pdf $\exp\Bigl\{-\bigl(\lambda-x_{(n)}+\xi^{(t)}\bigr)\Bigr\}$, for $\lambda>x_{(n)}-\xi^{(t)}$;
\item Compute $\eta$ as
\begin{equation*}
\eta=\min \left\{1,\frac{\pi\bigl(\lambda_{*}\big|\delta^{(t+1)}, \gamma^{(t+1)}, \xi^{(t)}, \boldsymbol{x}\bigr)\exp\bigl\{-\lambda_{(i-1)}\bigr\}}{\pi\bigl(\lambda_{(i-1)}\big|\delta^{(t+1)}, \gamma^{(t+1)}, \xi^{(t)}, \boldsymbol{x}\bigr) \exp\bigl\{-\lambda_{*}\bigr\}}\right\}.
\end{equation*}
\item Generate an uniform random variable on $(0,1)$, say $u$. If $u<\eta$, then $\lambda_{(i)}=\lambda_{*}$, otherwise $\lambda_{(i)}=\lambda_{(i-1)}$;
\item If $i=N$, then go to the next step. Otherwise set $i=i+1$ and go to step 2;
\item Accept $\lambda_{(N)}$ as a generation form $\pi\bigl(\lambda\big|\delta^{(t)}, \gamma^{(t)}, \xi^{(t)}, \boldsymbol{x}\bigr)$, i.e., $\lambda^{(t+1)}=\lambda_{(N)}$ and stop the MH algorithm. 
\end{enumerate}
\section{}\label{apd}
Similar to the $\lambda$, the full conditional pdf of $\xi$ has complicated structure. We have
\begin{align*}
\pi(\xi|\delta, \gamma, \lambda, \boldsymbol{x}) &\propto \Pi_{i=1}^{n}\Biggl(\frac{(\lambda+\xi-x_i)^{\delta \gamma-1}}{(x_i-\xi)^{\delta \gamma+1}}\Biggr)
\exp\left\{-\frac{\delta^2}{2}\sum_{i=1}^{n}\Bigg[\log\biggl(\frac{x_i-\xi}{\lambda+\xi-x_i}\biggr)\Bigg]^2\right\}\nonumber,
\end{align*}
where $x_{(n)}-\lambda<\xi<x_{(1)}$ in which $x_{(1)}=\min\{x_1,\dots,x_n\}$. Since the full conditional pdf of $\xi$ is not always log-concave, so we use the MH algorithm to sample from it. For this aim, we choose the uniform distribution on $\bigl(x_{(n)}-\lambda, x_{(1)}\bigr)$ as the proposal. The six-step MH algorithm is given by the following.
\begin{enumerate}
\item Suppose we are currently at $t$-th iteration of the sampler. Set $ i=1$ and generate a random variable from uniform distribution on $\bigl(x_{(n)}-\lambda^{(t+1)}, x_{(1)}\bigr)$, say $u$. Set $u=\xi^{(0)}$;
\item Sample $\xi^{*}$ from proposal distribution with pdf $q\bigl(\xi)=1/\bigl(x_{(1)}-x_{(n)}+\lambda^{(t+1)}\bigr)$;
\item Compute $\eta$ as
\begin{equation*}
\eta=\min \left\{1,\frac{\pi\bigl(\xi^{*}\big|\delta^{(t+1)}, \gamma^{(t+1)}, \lambda^{(t+1)}, \boldsymbol{x}\bigr)}{\pi\bigl(\xi^{(i-1)}\big|\delta^{(t+1)}, \gamma^{(t+1)}, \lambda^{(t+1)}, \boldsymbol{x}\bigr)}\right\}.
\end{equation*}
\item Generate an uniform random variable, say $u$, on $(0,1)$. If $u<\eta$, then $\xi^{(i)}=\xi^{*}$, otherwise $\xi^{(i)}=\xi^{(i-1)}$;
\item If $i=N$, then go to the next step. Otherwise set $i=i+1$ and go to step 2;
\item Accept $\xi_{(N)}$ as a generation form pdf $\pi\Bigl(\xi\big|\delta^{(t+1)}, \gamma^{(t+1)}, \lambda^{(t+1)}, \boldsymbol{x}\Bigr)$, i.e., $\xi^{(t+1)}=\xi_{(N)}$ and stop the MH algorithm. 
\end{enumerate}


%
\begin{table}[!h]
\caption{Percentage of runs that NR method truly converged for the JSB distribution.}
\begin{center} 
\begin{tabular}{cccccccc} 
\cline{1-8}
Sample size &            20 &           50&          100&          250&           500&        1000&       5000\\ \cline{1-8} 
Percentage  &      68.5\%&    68.7\%&     68.3\%&     68.3\%&     68.1\%&     68.4\%&    67.9\% \\ \cline{1-8}
\end{tabular} 
\end{center} 
\label{tab1}
\end{table}
\begin{table}[!h]
\caption{Summary statistics for DBH data.} 
\begin{center} 
\begin{tabular}{cccccccccc} 
\cline{1-10} 
  Plot &      Plot size&       Min&  1st Quar.&     Median&   Mean&   3rd Quar.&        Max.&   St.Dev.& Skewness\\ \cline{1-10} 
          9&                   52&    10.4&          17.32&       27.55&   28.01&            36.42&       55.9&     12.37&         0.48\\ \cline{1-10}
          44&                   42&    11.9&          14.78&       17.40&   24.27&            26.50&       83.8&     15.06&         2.13\\ \cline{1-10}
          73&                   35&     9.1&           13.85&       21.10&   30.35&            39.75&       88.6&     20.94&         1.15\\ \cline{1-10}
\end{tabular} 
\end{center} 
\label{tab2}
\end{table}
\begin{table}[!h]
\caption{Descriptive statistics for the sampler output.} 
\begin{tabular}{ccccccccc} 
\cline{1-9} 
Parameter& Min&  1st Quar.&     Median&   Mean&   3rd Quar.&  Max.&   St.Dev.& Skewness\\ \cline{1-9} 
 $\delta$   &     1.242&   1.724&   1.850&   1.874&  1.996&  15.000& 0.3406& 20.123\\ 
 $\gamma$& -15.000&   1.407&   1.615&   1.609&  1.804&  15.982&  0.450&   7.225\\
 $\lambda$&   12.399& 15.941&   17.215& 17.155& 18.410& 60.000&1.719&1.469\\
 $\xi$        &  -10.000&  -0.1270&  0.245&   0.160&   0.549&   1.399&0.542&-1.460\\
\cline{1-9}
\end{tabular}
\label{tab3}
\end{table}
\begin{table}[!h]
\caption{Bayesian estimators for the parameters of the JBS and three-parameter Weibull distributions.} 
\centering{\begin{tabular}{ccccccccc} 
\cline{1-9} 
           & \multicolumn{4}{c}{{{JSB distribution}}} &&  \multicolumn{3}{c}{Weibull distribution} \\ \cline{1-9} 
Plot&${\delta}$& ${\gamma}$ &${\lambda}$&$\xi$              &&$\alpha$& $\beta$ &$\mu$\\ \cline{1-9} 
 9 &0.772&0.545&52.311&8.719&&1.682&23.120&7.436\\
 44 &0.875&0.203&64.162&3.642&&2.145&34.496&2.171\\
 73 &0.641&0.978&88.592&8.182&&1.005&22.746&8.278\\
\cline{1-9} 
\end{tabular}}
\label{tab4}
\end{table}
\begin{table}[!h]
\caption{Computed goodness-of-fit statistics for modelling DBH data using JBS and three-parameter Weibull distributions.} 
\centering{\begin{tabular}{cccccccccc} 
\cline{1-10} 
           & \multicolumn{4}{c}{{{JSB distribution}}} &&  \multicolumn{4}{c}{Weibull distribution} \\ \cline{1-10} 
Plot&AD&CM&KS&LL              &&AD&CM&KS&LL\\ \cline{1-10} 
 9 &0.167&0.023&0.059&-196.808&&0.310&0.048&0.082&-199.579\\
 44 &0.216&0.028&0.100&-120.072&&0.302&0.040&0.115&-121.846\\
 73 &0.105&0.031&0.075&-141.751&&0.257&0.045&0.084&-143.299\\
\cline{1-10} 
\end{tabular}}
\label{tab5}
\end{table}

\end{document}